\documentclass[12pt]{article}
\usepackage{amsmath}
\usepackage{txfonts,bm,pifont}  
\usepackage[dvipdfm]{graphicx}
\usepackage[dvipdfm,bookmarks=true,bookmarksnumbered=true,colorlinks=true]{hyperref}
\newtheorem{theorem}{Theorem}[section]
\newtheorem{proposition}[theorem]{Proposition}
\newtheorem{corollary}[theorem]{Corollary}

\newtheorem{remark}[theorem]{Remark}
\newtheorem{definition}[theorem]{Definition}
\newcommand{\qed}{\qquad$\square$}


\def\disp{\displaystyle}
\def\aa{\alpha}

\def\dl{\delta}

\def\ep{\varepsilon}

\def\hf{\frac{1}{2}}

\def\bs{\backslash}
\newcommand{\BZ}{\mathbb Z}
\newcommand{\BC}{\mathbb C}

\newtheorem{thm}{Theorem}[section]

\newtheorem{prop}[thm]{Proposition}
\newtheorem{lem}[thm]{Lemma}

\makeatletter
 \renewcommand{\theequation}{%
  \thesection.\arabic{equation}}
 \@addtoreset{equation}{section}
\makeatother

\begin{document}

\title{An Explicit Formula for the Discrete Power Function Associated with Circle Patterns of Schramm Type}
\author{Hisashi Ando${}^1$, Mike Hay${}^2$, Kenji Kajiwara${}^2$ and Tetsu Masuda${}^3$,\\[2mm]
${}^1$Graduate School of Mathematics, Kyushu University, \\744 Motooka, Fukuoka 819-0395, Japan\\[1mm]
${}^2$Institute of Mathematics for Industry, Kyushu University,\\ 744 Motooka, Fukuoka 819-0395, Japan\\[1mm]
${}^3$Department of Physics and Mathematics, Aoyama Gakuin University,\\
Sagamihara, Kanagawa 229-8558, Japan}

\date{\today}
\maketitle
\begin{abstract}
We present an explicit formula for the discrete power function introduced by Bobenko, which is
expressed in terms of the hypergeometric $\tau$ functions for the sixth Painlev\'e equation. The
original definition of the discrete power function imposes strict conditions on the domain and the
value of the exponent. However, we show that one can extend the value of the exponent to arbitrary
complex numbers except even integers and the domain to a discrete analogue of the Riemann
surface. Moreover, we show that the discrete power function is an immersion when the real part of
the exponent is equal to one.
\end{abstract}

\section{Introduction}
The theory of discrete analytic functions has been developed in recent years based on the theory of
circle packings or circle patterns, which was initiated by Thurston's idea of using circle packings
as an approximation of the Riemann mapping\,\cite{Thurston}. So far many important properties have
been established for discrete analytic functions, such as the discrete maximum principle and
Schwarz's lemma\,\cite{Beardon-Stephenson}, the discrete uniformization theorem\,\cite{Rodin}, and
so forth. For a comprehensive introduction to the theory of discrete analytic functions, we refer to
\cite{Stephenson:book}.

It is known that certain circle patterns with fixed regular combinatorics admit rich structure. For
example, it has been pointed out that the circle patterns with square grid combinatorics introduced
by Schramm\,\cite{Schramm} and the hexagonal circle
patterns\,\cite{Agafonov-Bobenko:2003,Bobenko-Hoffmann,Bobenko-Hoffmann-Suris} are related to
integrable systems. Some explicit examples of discrete analogues of analytic functions have been
presented which are associated with Schramm's patterns: $\exp(z)$, ${\rm erf}(z)$, Airy
function\,\cite{Schramm}, $z^\gamma$, $\log(z)$\,\cite{Agafonov-Bobenko:2000}. Also, discrete
analogues of $z^\gamma$ and $\log(z)$ associated with hexagonal circle patterns are discussed in
\cite{Agafonov-Bobenko:2003,Bobenko-Hoffmann,Bobenko-Hoffmann-Suris}.

Among those examples, it is remarkable that the discrete analogue of the power function $z^\gamma$
associated with the circle patterns of Schramm type has a close relationship with the sixth
Painlev\'e equation (P$_{\rm VI}$)\,\cite{Bobenko}. It is desirable to construct a representation
formula for the discrete power function in terms of the Painlev\'e transcendents as was mentioned in
\cite{Bobenko}. The discrete power function can be formulated as a solution to a system of
difference equations on the square lattice $(n,m)\in\mathbb{Z}^2$ with a certain initial
condition. A correspondence between the dependent variable of this system and the Painlev\'e
transcendents can be found in \cite{NRGO}, but the formula seems somewhat indirect. Agafonov has
constructed a formula for the radii of circles of the associated circle pattern at some special
points on $\mathbb{Z}^2$ in terms of the Gauss hypergeometric function\,\cite{Agafonov_2}.  In this
paper, we aim to establish an explicit representation formula of the discrete power function itself
in terms of the hypergeometric $\tau$ function of P$_{\rm VI}$ which is valid on
$\mathbb{Z}^2_+=\{(n,m)\in\mathbb{Z}^2\,|\,n,m\ge0\}$ and for
$\gamma\in\mathbb{C}\backslash2\mathbb{Z}$. Based on this formula, we generalize the domain of the
discrete power function to a discrete analogue of the Riemann surface. 

On the other hand, the fact that the discrete power function is related to P$_{\rm VI}$ has been
used to establish the immersion property\,\cite{Agafonov-Bobenko:2000} and
embeddedness\,\cite{Agafonov_1} of the discrete power function with real exponent. Although we
cannot expect such properties and thus the correspondence to a certain circle pattern for general
complex exponent, we have found a special case of ${\rm Re}\,\gamma=1$ where the discrete power
function is an immersion. Another purpose of this paper is to prove the immersion property of this
case.

This paper is organized as follows. In section 2, we give a brief review of the definition of the
discrete power function and its relation to P$_{\rm VI}$. The explicit formula for the discrete
power function is given in section 3. We discuss the extension of the domain of the discrete power
function in section 4. In section 5, we show that the discrete power function for ${\rm
Re}~\gamma=1$ is an immersion. Section 6 is devoted to concluding remarks.

\section{Discrete power function}
\subsection{Definition of the discrete power function}
For maps, a discrete analogue of conformality has been proposed by Bobenko and Pinkall in the
framework of discrete differential geometry\,\cite{BP_1}.

\begin{definition}\label{d-conformal}
A map $f\,:\,\mathbb{Z}^2\to\mathbb{C}\,;\,(n,m)\mapsto f_{n,m}$ is called discrete conformal if the
cross-ratio with respect to every elementary quadrilateral is equal to $-1$:
\begin{equation}
\dfrac{(f_{n,m}-f_{n+1,m})(f_{n+1,m+1}-f_{n,m+1})}
{(f_{n+1,m}-f_{n+1,m+1})(f_{n,m+1}-f_{n,m})}=-1.
\label{cr}
\end{equation}
\end{definition}

The condition (\ref{cr}) is a discrete analogue of the Cauchy-Riemann relation. Actually, a smooth
map $f:D\subset\mathbb{C}\to\mathbb{C}$ is conformal if and only if it satisfies
\begin{equation}
\lim_{\epsilon\to0}
\dfrac{(f(x,y)-f(x+\epsilon,y))(f(x+\epsilon,y+\epsilon)-f(x,y+\epsilon))}
{(f(x+\epsilon,y)-f(x+\epsilon,y+\epsilon))(f(x,y+\epsilon)-f(x,y))}=-1
\label{CR}
\end{equation}
for all $(x,y)\in D$. However, using Definition \ref{d-conformal} alone, one cannot exclude maps
whose behaviour is far from that of usual holomorphic maps. Because of this, an additional condition
for a discrete conformal map has been
considered\,\cite{Agafonov_1,Agafonov-Bobenko:2000,Bobenko,BP_2}.

\begin{definition}
A discrete conformal map $f_{n,m}$ is called embedded if inner parts of different elementary
quadrilaterals $(f_{n,m},f_{n+1,m},f_{n+1,m+1},f_{n,m+1})$ do not intersect.
\end{definition}

\begin{figure}
\centering
\begin{tabular}{cc}
\begin{minipage}{0.4\textwidth}
\begin{center}
\includegraphics[scale=0.3]{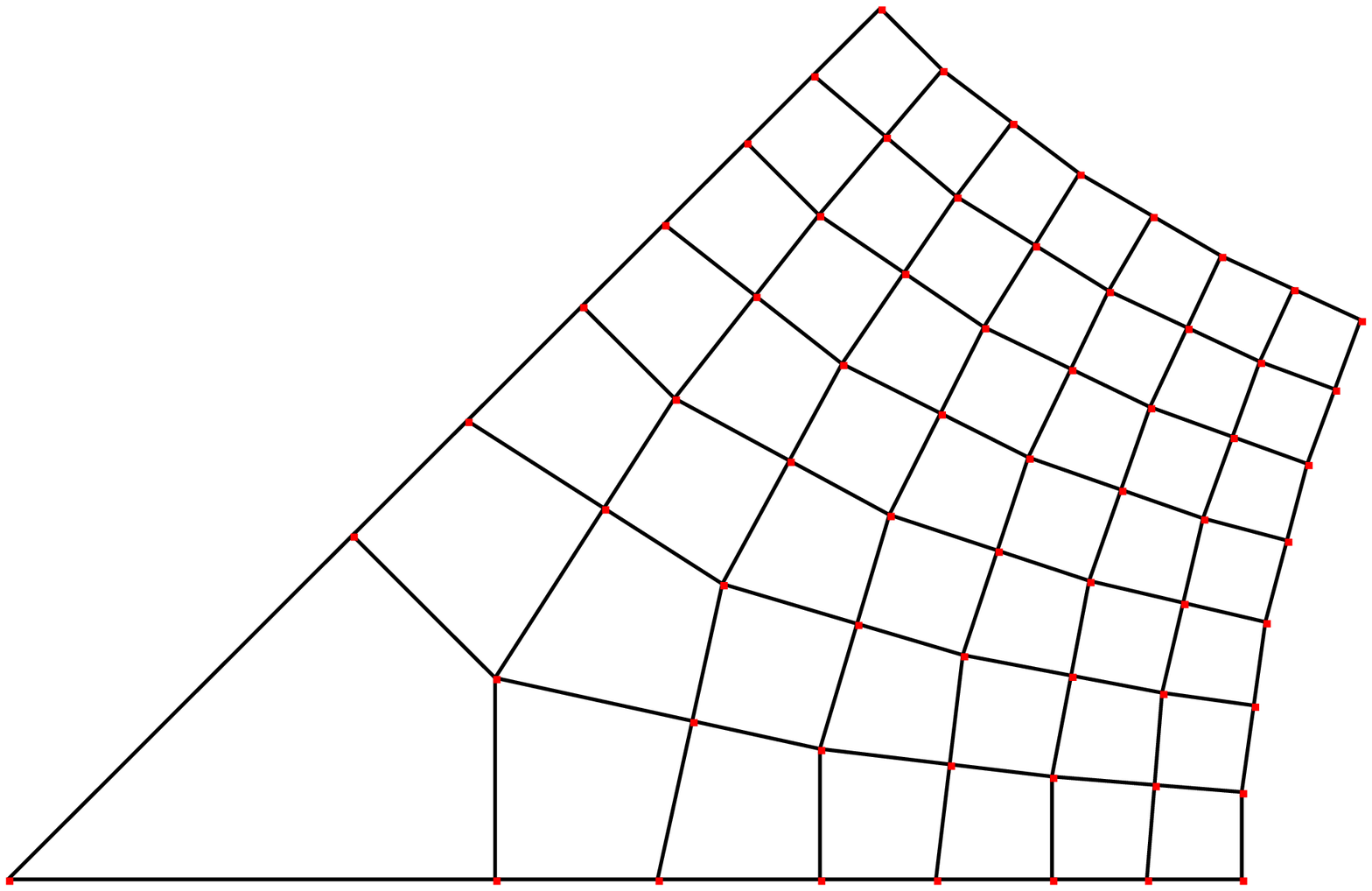}
\caption{\small{An example of the embedded discrete conformal map.}}
\label{embedded}
\end{center}
\end{minipage}
&
\begin{minipage}{0.4\textwidth}
\begin{center}
\includegraphics[scale=0.2]{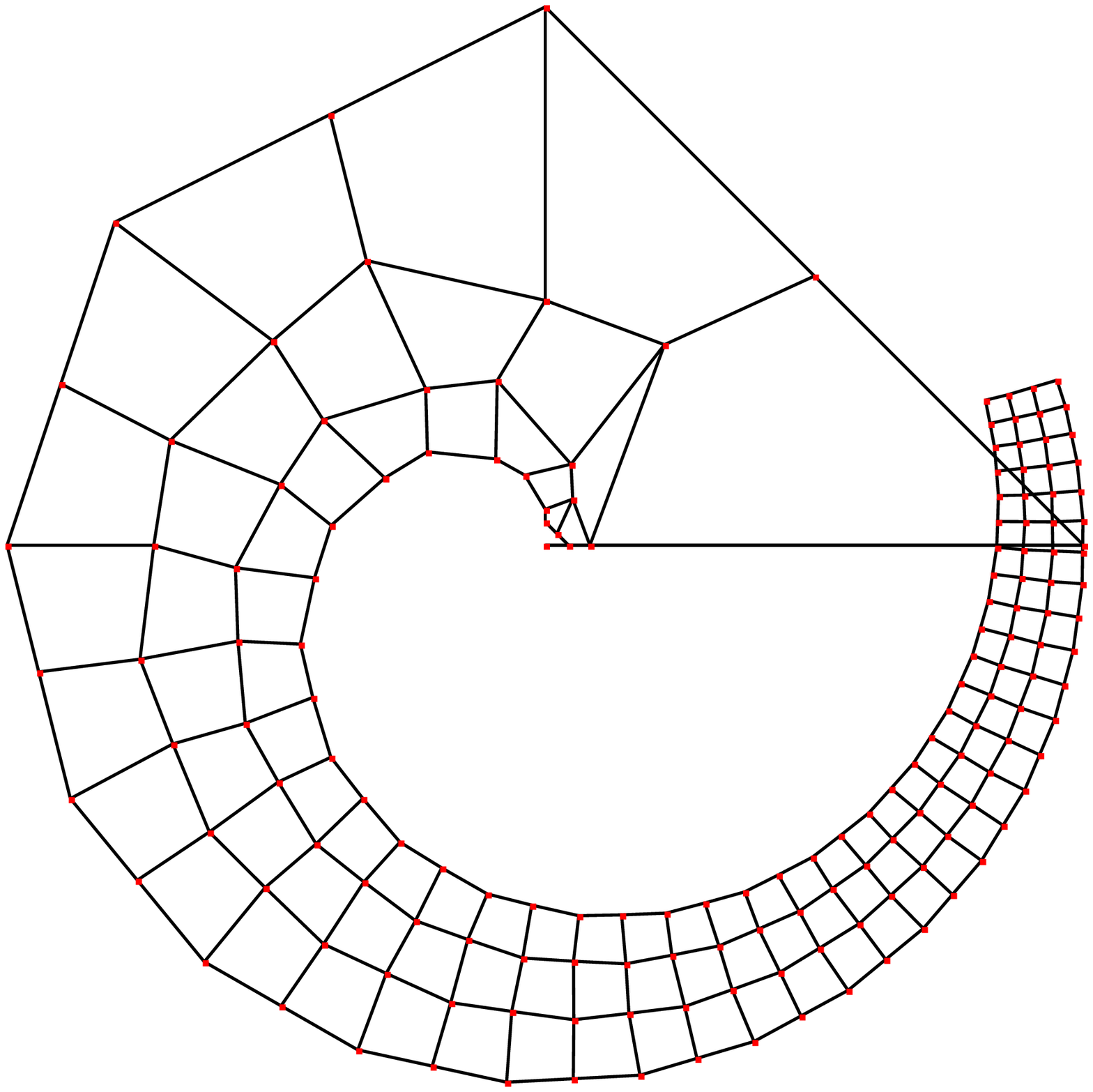}
\caption{\small{An example of the discrete conformal map that is not embedded but immersed.}}
\label{immersed}
\end{center}
\end{minipage}
\end{tabular}
\end{figure}

An example of an embedded map is presented in Figure \ref{embedded}. This condition seems to require
that $f=f_{n,m}$ is a univalent function in the continuous limit, and is too strict to capture a
wide class of discrete holomorphic functions. In fact, a relaxed requirement has been considered as
follows\,\cite{Agafonov_1,Agafonov-Bobenko:2000}.

\begin{definition}
A discrete conformal map $f_{n,m}$ is called immersed, or an immersion, if inner parts of adjacent
elementary quadrilaterals $(f_{n,m},f_{n+1,m},f_{n+1,m+1},f_{n,m+1})$ are disjoint.
\end{definition}

See Figure \ref{immersed} for an example of an immersed map. 

Let us give the definition of the discrete power function proposed by
Bobenko\,\cite{Agafonov-Bobenko:2000,Bobenko,BP_2}.

\begin{definition} \label{def:dpower1}
Let $f\,:\mathbb{Z}_+^2\to\mathbb{C}\,;\,(n,m)\mapsto f_{n,m}$ be a discrete conformal map. If
$f_{n,m}$ is the solution to the difference equation
\begin{equation}
\gamma f_{n,m}
=2n\dfrac{(f_{n+1,m}-f_{n,m})(f_{n,m}-f_{n-1,m})}{f_{n+1,m}-f_{n-1,m}}
+2m\dfrac{(f_{n,m+1}-f_{n,m})(f_{n,m}-f_{n,m-1})}{f_{n,m+1}-f_{n,m-1}}
\label{eq-f}
\end{equation}
with the initial conditions 
\begin{equation}
f_{0,0}=0,\quad f_{1,0}=1,\quad f_{0,1}=e^{\gamma\pi i/2}
\end{equation}
for $0<\gamma<2$, then we call $f$ a discrete power function. 
\end{definition}

The difference equation (\ref{eq-f}) is a discrete analogue of the differential equation $\gamma
f=z\dfrac{\partial f}{\partial z}$ for the power function $f(z)=z^{\gamma}$, which means that the
parameter $\gamma$ corresponds to the exponent of the discrete power function.

It is easy to get the explicit formula of the discrete power function for $m=0$ (or $n=0$). When
$m=0$, (\ref{eq-f}) is reduced to a three-term recurrence relation. Solving it with the initial
condition $f_{0,0}=0,\,f_{1,0}=1$, we have
\begin{equation}
f_{n,0}=\left\{
\begin{array}{ll}
\displaystyle
\dfrac{2l}{2l+\gamma}\prod_{k=1}^l\dfrac{2k+\gamma}{2k-\gamma}&(n=2l),\\[4mm]
\displaystyle\prod_{k=1}^l\dfrac{2k+\gamma}{2k-\gamma}&(n=2l+1),
\end{array}
\right.
\end{equation}
for $n\in\mathbb{Z}_+$. When $m=1$ (or $n=1$), Agafonov has shown that the discrete power function
can be expressed in terms of the hypergeometric function\,\cite{Agafonov_2}. One of the aims of this
paper is to give an explicit formula for the discrete power function $f_{n,m}$ for arbitrary
$(n,m)\in\mathbb{Z}_+^2$.

In Definition \ref{def:dpower1}, the domain of the discrete power function is restricted to the
``first quadrant'' $\mathbb{Z}_+^2$, and the exponent $\gamma$ to the interval $0<\gamma<2$. Under
this condition, it has been shown that the discrete power function is
embedded\,\cite{Agafonov_1}. For our purpose, we do not have to persist with such a restriction. In
fact, the explicit formula we will give is applicable to the case
$\gamma\in\mathbb{C}\backslash2\mathbb{Z}$. Regarding the domain, one can extend it to a discrete
analogue of the Riemann surface.

\subsection{Relationship to P$_{\rm VI}$\label{LSKdV-P6}}
In order to construct an explicit formula for the discrete power function $f_{n,m}$, we will move to
a more general setting. The cross-ratio condition (\ref{cr}) can be regarded as a special case of
the discrete Schwarzian KdV equation
\begin{equation}
\dfrac{(f_{n,m}-f_{n+1,m})(f_{n+1,m+1}-f_{n,m+1})}
{(f_{n+1,m}-f_{n+1,m+1})(f_{n,m+1}-f_{n,m})}=\dfrac{p_n}{q_m},
\end{equation}
where $p_n$ and $q_m$ are arbitrary functions in the indicated variables. Some of the authors have
constructed various special solutions to the above equation\,\cite{HKM}. In particular, they have
shown that an autonomous case
\begin{equation}
\dfrac{(f_{n,m}-f_{n+1,m})(f_{n+1,m+1}-f_{n,m+1})}
{(f_{n+1,m}-f_{n+1,m+1})(f_{n,m+1}-f_{n,m})}=\dfrac{1}{t},
\label{LS-KdV}
\end{equation}
where $t$ is independent of $n$ and $m$, can be regarded as a part of the B\"acklund transformations
of P$_{\rm VI}$, and given special solutions to (\ref{LS-KdV}) in terms of the $\tau$ functions of
P$_{\rm VI}$.

We here give a brief account of the derivation of P$_{\rm VI}$ according to \cite{NRGO}. The
derivation is achieved by imposing a certain similarity condition on the discrete Schwarzian KdV
equation (\ref{LS-KdV}) and the difference equation (\ref{eq-f}) simultaneously. The discrete
Schwarzian KdV equation (\ref{LS-KdV}) is automatically satisfied if there exists a function
$v_{n,m}$ satisfying
\begin{equation}
f_{n,m}-f_{n+1,m}=t^{-1/2}v_{n,m}v_{n+1,m},\quad
f_{n,m}-f_{n,m+1}=v_{n,m}v_{n,m+1}. 
\label{f-v}
\end{equation}
By eliminating the variable $f_{n,m}$, we get for $v_{n,m}$ the following equation
\begin{equation}
t^{1/2}v_{n,m}v_{n,m+1}+v_{n,m+1}v_{n+1,m+1}
=v_{n,m}v_{n+1,m}+t^{1/2}v_{n+1,m}v_{n+1,m+1},
\label{compatibilty}
\end{equation}
which is equivalent to the lattice modified KdV equation. It can be shown that the difference
equation (\ref{eq-f}) is reduced to
\begin{equation}
n\dfrac{v_{n+1,m}-v_{n-1,m}}{v_{n+1,m}+v_{n-1,m}}+
m\dfrac{v_{n,m+1}-v_{n,m-1}}{v_{n,m+1}+v_{n,m-1}}=\mu-(-1)^{m+n}\lambda
\label{eq-v}
\end{equation}
with $\gamma=1+2\mu$, where $\lambda\in\mathbb{C}$ is an integration constant. In the following we
take $\lambda=\mu$ so that (\ref{eq-v}) is consistent when $n=m=0$ and $v_{1,0}+v_{-1,0}\ne0\ne
v_{0,1}+v_{0,-1}$.

Assume that the dependence of the variable $v_{n,m}=v_{n,m}(t)$ on the deformation parameter $t$ is given by 
\begin{equation}
-2t\dfrac{d}{dt}\log v_{n,m}
=n\dfrac{v_{n+1,m}-v_{n-1,m}}{v_{n+1,m}+v_{n-1,m}}+\chi_{n+m},
\label{similarity}
\end{equation}
where $\chi_{n+m}=\chi_{n+m}(t)$ is an arbitrary function satisfying $\chi_{n+m+2}=\chi_{n+m}$. Then
we have the following Proposition.

\begin{proposition}
Let $q=q_{n,m}=q_{n,m}(t)$ be the function defined by
$q_{n,m}=t^{1/2}\dfrac{v_{n+1,m}}{v_{n,m+1}}$. Then $q$ satisfies P$_{\rm VI}$
\begin{equation}
\begin{array}{l}
\dfrac{d^2q}{dt^2}=
\dfrac{1}{2}\left(\dfrac{1}{q}+\dfrac{1}{q-1}+\dfrac{1}{q-t}\right)
\left(\dfrac{dq}{dt}\right)^2
-\left(\dfrac{1}{t}+\dfrac{1}{t-1}+\dfrac{1}{q-t}\right)\dfrac{dq}{dt}\\[4mm]
\hskip15mm
+\dfrac{q(q-1)(q-t)}{2t^2(t-1)^2}
\left[\kappa_{\infty}^2-\kappa_0^2\dfrac{t}{q^2}
     +\kappa_1^2\dfrac{t-1}{(q-1)^2}+(1-\theta^2)\dfrac{t(t-1)}{(q-t)^2}\right],
\end{array}   \label{P6}
\end{equation}
with
\begin{equation}
\begin{array}{ll}
\kappa_{\infty}^2=\dfrac{1}{4}(\mu-\nu+m-n)^2,&
\kappa_0^2=\dfrac{1}{4}(\mu-\nu-m+n)^2,\\[4mm]
\kappa_1^2=\dfrac{1}{4}(\mu+\nu-m-n-1)^2,&
\theta^2=\dfrac{1}{4}(\mu+\nu+m+n+1)^2,
\end{array}
\end{equation}
where we denote $\nu=(-1)^{m+n}\mu$.
\end{proposition}

In general, P$_{\rm VI}$ contains four complex parameters denoted by
$\kappa_{\infty},\kappa_0,\kappa_1$ and $\theta$. Since $n,m\in\mathbb{Z}_+$, a special case of
P$_{\rm VI}$ appears in the above proposition, which corresponds to the case where P$_{\rm VI}$
admits special solutions expressible in terms of the hypergeometric function. In fact, the special
solutions to P$_{\rm VI}$ of hypergeometric type are given as follows:

\begin{proposition}\label{HG_P6}
\,\cite{Masuda}\quad Define the function $\tau_{n'}(a,b,c;t)\,(c\notin\mathbb{Z},\,n'\in\mathbb{Z}_+)$ by 
\begin{equation}
\tau_{n'}(a,b,c;t)=\left\{
\begin{array}{cc}
\displaystyle\det\left(\varphi(a+i-1,b+j-1,c;t)\right)_{1\le i,j\le n'}
&(n'>0),\\
\displaystyle 1& (n'=0),
\end{array}\right.
\label{tau}
\end{equation}
with 
\begin{equation}
\begin{array}{l}
\varphi(a,b,c;t)
=c_0\dfrac{\Gamma(a)\Gamma(b)}{\Gamma(c)}F(a,b,c;t)\\[3mm]
\hskip22mm
+c_1\dfrac{\Gamma(a-c+1)\Gamma(b-c+1)}{\Gamma(2-c)}t^{1-c}F(a-c+1,b-c+1,2-c;t).
\end{array}   \label{Gauss}
\end{equation}
Here, $F(a,b,c;t)$ is the Gauss hypergeometric function, $\Gamma(x)$ is the Gamma function, and
$c_0$ and $c_1$ are arbitrary constants. Then
\begin{equation}
q=\dfrac{\tau_{n'}^{0,-1,0}\tau_{n'+1}^{-1,-1,-1}}
        {\tau_{n'}^{-1,-1,-1}\tau_{n'+1}^{0,-1,0}}
\end{equation}
with $\tau_{n'}^{k,l,m}=\tau_{n'}(a+k+1,b+l+2,c+m+1;t)$ gives a family of hypergeometric solutions
to P$_{\rm VI}$ with the parameters
\begin{equation}
\kappa_{\infty}=a+n',\quad\kappa_0=b-c+1+n',\quad\kappa_1=c-a,\quad\theta=-b.
\end{equation}
\end{proposition}

We call $\tau_{n'}(a,b,c;t)$ or $\tau_{n'}^{k,l,m}$ the hypergeometric $\tau$ function of P$_{\rm VI}$.

\section{Explicit formulae}
\subsection{Explicit formulae for $f_{n,m}$ and $v_{n,m}$}
We present the solution to the simultaneous system of the discrete Schwarzian KdV equation
(\ref{LS-KdV}) and the difference equation (\ref{eq-f}) under the initial conditions
\begin{equation}
f_{0,0}=0,\quad f_{1,0}=c_0,\quad f_{0,1}=c_1t^r,\label{initial}
\end{equation}
where $\gamma=2r$, and $c_0$ and $c_1$ are arbitrary constants. We set $c_0=c_1=1$ and 
$t=e^{\pi i}(=-1)$ to obtain the explicit formula for the original discrete power function. Note that
$\tau_{n'}(b,a,c;t)=\tau_{n'}(a,b,c;t)$ by the definition. Moreover, we interpret $F(k,b,c;t)$ for
$k\in\mathbb{Z}_{>0}$ as $F(k,b,c;t)=0$ and $\Gamma(-k)$ for $k\in\mathbb{Z}_{\ge0}$ as
$\Gamma(-k)=\dfrac{(-1)^k}{k!}$.

\begin{theorem}\label{formula_f}
For $(n,m)\in\mathbb{Z}_+^2$, the function $f_{n,m}=f_{n,m}(t)$ can be expressed as follows. 

\begin{enumerate}
\item Case where $n\le m$ (or $n'=n$). When $n+m$ is even, we have 
\begin{equation}
f_{n,m}=c_1t^{r-n}N\dfrac{(r+1)_{N-1}}{(-r+1)_N}
\dfrac{\tau_n(-N,-r-N+1,-r;t)}{\tau_n(-N+1,-r-N+2,-r+2;t)},
\end{equation}
where $N=\dfrac{n+m}{2}$ and $(u)_j=u(u+1)\cdots(u+j-1)$ is the Pochhammer symbol. When $n+m$ is odd, we have 
\begin{equation}
f_{n,m}=c_1t^{r-n}\dfrac{(r+1)_{N-1}}{(-r+1)_{N-1}}
\dfrac{\tau_n(-N+1,-r-N+1,-r;t)}{\tau_n(-N+2,-r-N+2,-r+2;t)},
\end{equation}
where $N=\dfrac{n+m+1}{2}$. 

\item Case where $n\ge m$ (or $n'=m$). When $n+m$ is even, we have 
\begin{equation}
f_{n,m}=c_0N\dfrac{(r+1)_{N-1}}{(-r+1)_N}
\dfrac{\tau_m(-N+2,-r-N+1,-r+2;t)}{\tau_m(-N+1,-r-N+2,-r+2;t)},
\end{equation}
where $N=\dfrac{n+m}{2}$. When $n+m$ is odd, we have 
\begin{equation}
f_{n,m}=c_0\dfrac{(r+1)_{N-1}}{(-r+1)_{N-1}}
\dfrac{\tau_m(-N+2,-r-N+1,-r+1;t)}{\tau_m(-N+1,-r-N+2,-r+1;t)},
\end{equation}
where $N=\dfrac{n+m+1}{2}$.
\end{enumerate}
\end{theorem}

\begin{proposition}\label{formula_v}
For $(n,m)\in\mathbb{Z}_+^2$, the function $v_{n,m}=v_{n,m}(t)$ can be expressed as follows. 

\begin{enumerate}
\item Case where $n\le m$ (or $n'=n$). When $n+m$ is even, we have 
\begin{equation}
v_{n,m}=t^{-\frac{n}{2}}\dfrac{(r)_N}{(-r+1)_N}
\dfrac{\tau_{n}(-N+1,-r-N+1,-r+1;t)}{\tau_{n}(-N+1,-r-N+2,-r+2;t)},
\end{equation}
where $N=\dfrac{n+m}{2}$. When $n+m$ is odd, we have 
\begin{equation}
v_{n,m}=-c_1t^{r-\frac{n}{2}}
\dfrac{\tau_{n}(-N+1,-r-N+2,-r+1;t)}{\tau_{n}(-N+2,-r-N+2,-r+2;t)},
\end{equation}
where $N=\dfrac{n+m+1}{2}$. 

\item Case where $n\ge m$ (or $n'=m$). When $n+m$ is even, we have 
\begin{equation}
v_{n,m}=t^{-\frac{m}{2}}\dfrac{(r)_N}{(-r+1)_N}
\dfrac{\tau_m(-N+1,-r-N+1,-r+1;t)}{\tau_m(-N+1,-r-N+2,-r+2;t)},
\end{equation}
where $N=\dfrac{n+m}{2}$. When $n+m$ is odd, we have 
\begin{equation}
\begin{array}{l}
v_{n,m}=-c_0t^{\frac{m+1}{2}}
\dfrac{\tau_m(-N+2,-r-N+2,-r+2;t)}{\tau_m(-N+1,-r-N+2,-r+1;t)},
\end{array}
\end{equation}
where $N=\dfrac{n+m+1}{2}$. 
\end{enumerate}
\end{proposition}

\begin{figure}
\centering
\begin{tabular}{cc}
\begin{minipage}{0.4\textwidth}
\begin{center}
\includegraphics[scale=0.2]{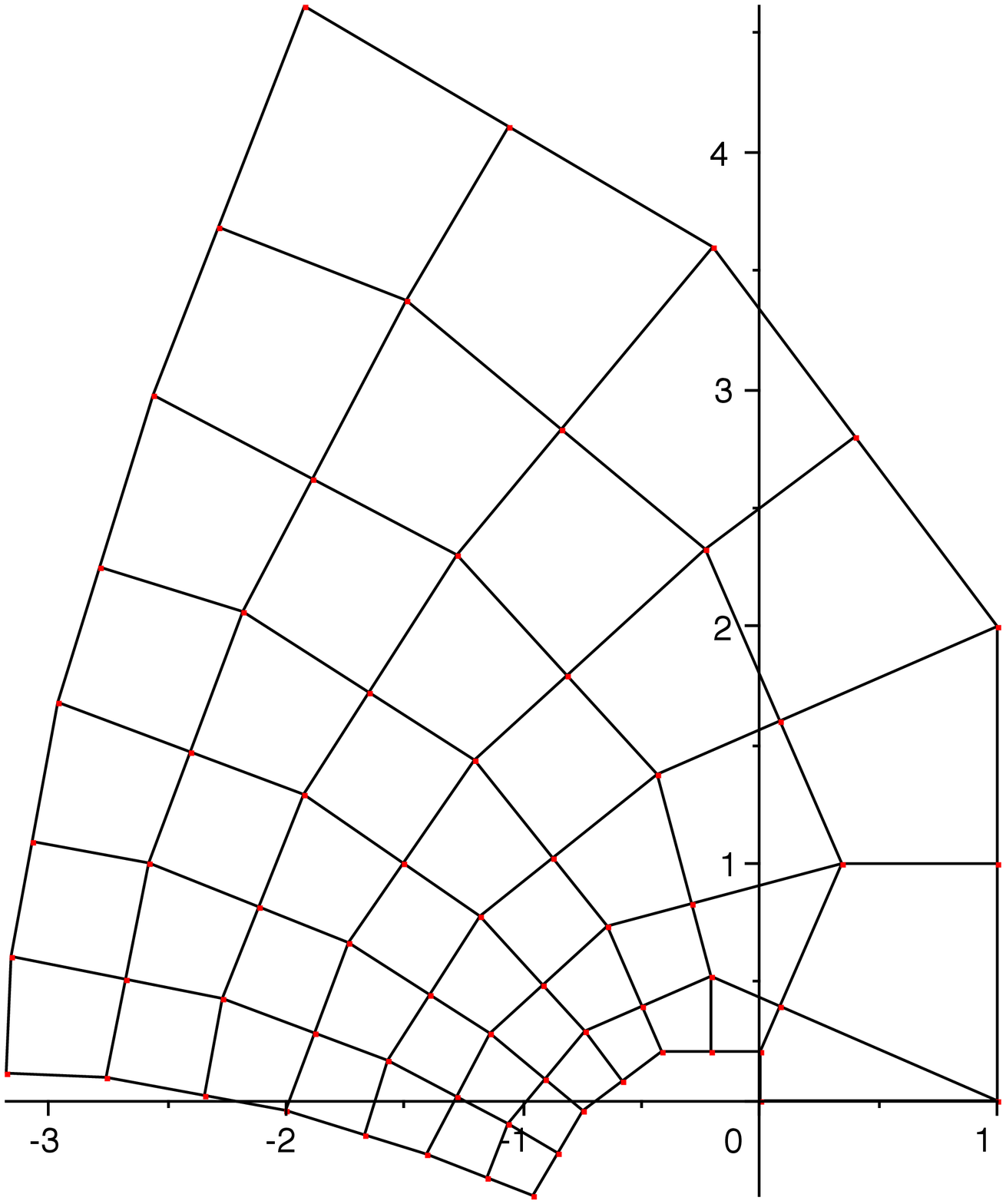}
\caption{{\small The discrete power function with $\gamma=1+i$.}}
\label{1+i}
\end{center}
\end{minipage}
&
\begin{minipage}{0.4\textwidth}
\begin{center}
\includegraphics[scale=0.2]{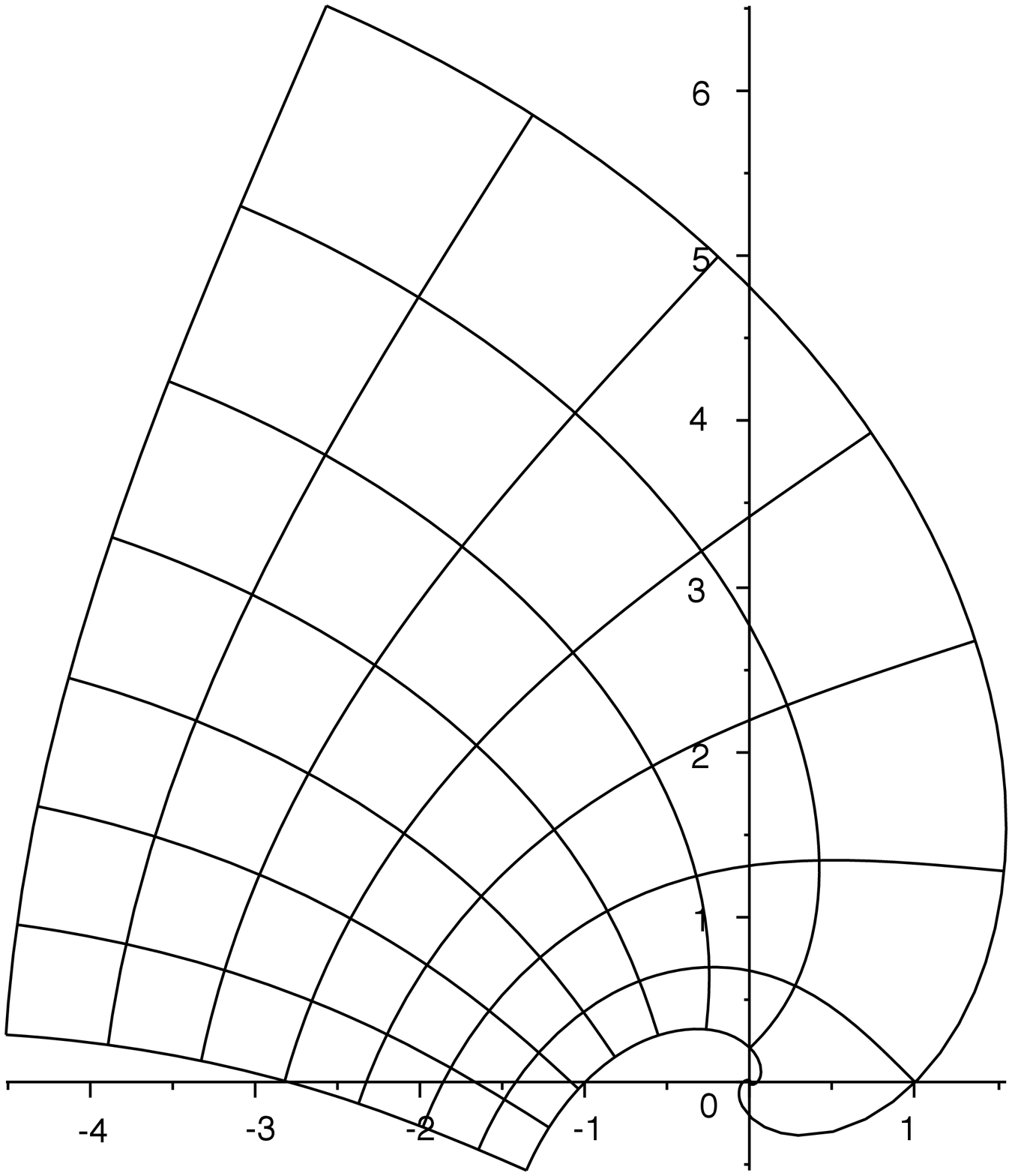}
\caption{{\small The ordinary power function $z^{1+i}$.}}
\label{continuous}
\end{center}
\end{minipage}
\end{tabular}
\end{figure}

\begin{figure}
\centering
\begin{minipage}{0.4\textwidth}
\includegraphics[scale=0.3]{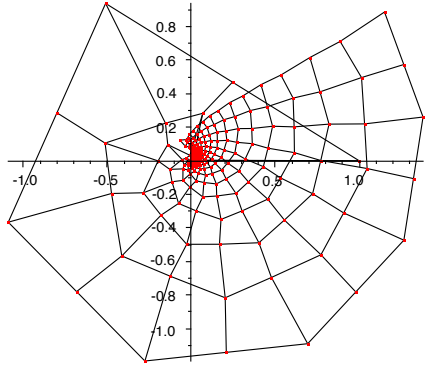}
\caption{{\small The discrete power function with $\gamma=0.25+3.35i$.}}
\label{0.25+3.35i}
\end{minipage}
\end{figure}

Note that these expressions are applicable to the case where $r\in\mathbb{C}\backslash\mathbb{Z}$. A
typical example of the discrete power function and its continuous counterpart are illustrated in
Figure \ref{1+i} and Figure \ref{continuous}, respectively. Figure \ref{0.25+3.35i} shows an example
of the case suggesting multivalency of the map. The
proof of the above theorem and proposition is given in the next subsection.

\begin{remark}
Agafonov has shown that the generalized discrete power function $f_{n,m}$, under the setting of $c_0=c_1=1$, $t=e^{2i\alpha}\,(0<\alpha<\pi)$ and $0<r<1$, is embedded\,\cite{Agafonov_2}.
\end{remark}

\begin{remark}
As we mention above, some special solutions to (\ref{LS-KdV}) in terms of the $\tau$ functions of
P$_{\rm VI}$ have been presented\,\cite{HKM}. It is easy to show that these solutions also satisfy a
difference equation which is a deformation of (\ref{eq-f}) in the sense that the coefficients $n$
and $m$ of (\ref{eq-f}) are replaced by arbitrary complex numbers. For instance, a class of
solutions presented in Theorem 6 of \cite{HKM} satisfies
\begin{equation}
\begin{array}{l}
(\alpha_0+\alpha_2+\alpha_4)f_{n,m}\\[1mm]
=(n-\alpha_2)
\dfrac{(f_{n+1,m}-f_{n,m})(f_{n,m}-f_{n-1,m})}{f_{n+1,m}-f_{n-1,m}}
-(\alpha_1+\alpha_2+\alpha_4-m)
\dfrac{(f_{n,m+1}-f_{n,m})(f_{n,m}-f_{n,m-1})}{f_{n,m+1}-f_{n,m-1}},
\end{array}
\end{equation}
where $\alpha_i$ are parameters of P$_{\rm VI}$ introduced in Appendix \ref{sym_P6}. Setting the
parameters as $(\alpha_0,\alpha_1,\alpha_2,\alpha_3,\alpha_4)=(r,0,0,-r+1,0)$, we see that the above
equation is reduced to (\ref{eq-f}) and that the solutions are given by the hypergeometric $\tau$
functions under the initial conditions (\ref{initial}).
\end{remark}

\subsection{Proof of the results}
In this subsection, we give the proof of Theorem \ref{formula_f} and Proposition
\ref{formula_v}. One can easily verify that $f_{n,m}$ satisfies the initial condition
(\ref{initial}) by noticing $\tau_0(a,b,c;t)=1$. We then show that $f_{n,m}$ and $v_{n,m}$ given in
Theorem \ref{formula_f} and Proposition \ref{formula_v} satisfy the relation (\ref{f-v}), the
difference equation (\ref{eq-f}), the compatibility condition (\ref{compatibilty}) and the
similarity condition (\ref{similarity}) by means of the various bilinear relations for the
hypergeometric $\tau$ function. Note in advance that we use the bilinear relations by specializing
the parameters $a,b$ and $c$ as
\begin{equation}
a=-N,\quad b=-r-N,\quad c=-r+1,\quad N=\frac{n+m}{2},\label{para_even}
\end{equation}
when $n+m$ is even, or 
\begin{equation}
a=-r-N+1,\quad b=-N,\quad c=-r+1,\quad N=\frac{n+m+1}{2},\label{para_odd}
\end{equation}
when $n+m$ is odd. 

We first verify the relation (\ref{f-v}). Note that we have the following bilinear relations 
\begin{equation}
\begin{array}{l}
(c-1)\tau_n^{0,-1,-1}\tau_{n+1}^{-1,-1,-1}
=(c-b-1)t\tau_{n+1}^{0,-1,0}\tau_n^{-1,-1,-2}
+b\tau_n^{0,0,0}\tau_{n+1}^{-1,-2,-2},\\[2mm]
(c-1)\tau_n^{-1,-1,-1}\tau_n^{0,-1,-1}
=(c-b-1)\tau_n^{0,-1,0}\tau_n^{-1,-1,-2}
+b\tau_n^{0,0,0}\tau_n^{-1,-2,-2},
\end{array}   \label{bi_n}
\end{equation}
\begin{equation}
\begin{array}{l}
(a-b)\tau_m^{0,-1,-1}\tau_m^{0,-1,0}
=a\tau_m^{-1,-1,-1}\tau_m^{1,-1,0}-b\tau_m^{0,0,0}\tau_m^{0,-2,-1},\\[2mm]
(a-b)t\tau_{m+1}^{0,-1,0}\tau_m^{0,-1,-1}
=a\tau_{m+1}^{-1,-1,-1}\tau_m^{1,-1,0}-b\tau_m^{0,0,0}\tau_{m+1}^{0,-2,-1},
\end{array}   \label{bi_m}
\end{equation}
\begin{equation}
\begin{array}{l}
(b-a+1)\tau_m^{0,0,0}\tau_m^{-1,-1,-1}
=(b-c+1)\tau_m^{0,-1,0}\tau_m^{-1,0,-1}
+(c-a)\tau_m^{0,-1,-1}\tau_m^{-1,0,0},\\[2mm]
(b-a+1)\tau_{m+1}^{-1,-1,-1}\tau_m^{0,0,0}
=(b-c+1)\tau_{m+1}^{0,-1,0}\tau_m^{-1,0,-1}
+(c-a)\tau_m^{0,-1,-1}\tau_{m+1}^{-1,0,0},
\end{array}   \label{bi_m'}
\end{equation}
for the hypergeometric $\tau$ functions, the derivation of which is discussed in Appendix
\ref{sym_P6}. Let us consider the case where $n'=n$. When $n+m$ is even, the relation (\ref{f-v}) is
reduced to
\begin{equation}
\begin{array}{l}
-r\tau_{n}^{[1,1,1]}\tau_{n+1}^{[0,1,1]}
=Nt\tau_{n+1}^{[1,1,2]}\tau_{n}^{[0,1,0]}
-(r+N)\tau_{n}^{[1,2,2]}\tau_{n+1}^{[0,0,0]},\\[1mm]
-r\tau_{n}^{[0,1,1]}\tau_{n}^{[1,1,1]}
=N\tau_{n}^{[1,1,2]}\tau_{n}^{[0,1,0]}
-(r+N)\tau_{n}^{[1,2,2]}\tau_{n}^{[0,0,0]},
\end{array}   \label{bi:f-v}
\end{equation}
where we denote 
\begin{equation}
\tau_{n'}^{[i_1,i_2,i_3]}=\tau_{n'}(-N+i_1,-r-N+i_2,-r+i_3;t),
\end{equation}
for simplicity. We see that the relations (\ref{bi:f-v}) can be obtained from (\ref{bi_n}) with the
parameters specialized as (\ref{para_even}). In fact, the hypergeometric $\tau$ functions can be
rewritten as
\begin{equation}
\tau_n^{0,-1,-1}=\tau_n(a+1,b+1,c)=\tau_n(-N+1,-r-N+1,-r+1)=\tau_n^{[1,1,1]},
\end{equation}
for instance. When $n+m$ is odd, (\ref{f-v}) yields
\begin{equation}
\begin{array}{l}
-r\tau_{n}^{[1,2,1]}\tau_{n+1}^{[1,1,1]}
=(-r+N)t\tau_{n+1}^{[1,2,2]}\tau_{n}^{[1,1,0]}
-N\tau_{n}^{[2,2,2]}\tau_{n+1}^{[0,1,0]},\\[1mm]
-r\tau_{n}^{[1,1,1]}\tau_{n}^{[1,2,1]}
=(-r+N)\tau_{n}^{[1,2,2]}\tau_{n}^{[1,1,0]}
-N\tau_{n}^{[2,2,2]}\tau_{n}^{[0,1,0]},
\end{array}
\end{equation}
which is also obtained from (\ref{bi_n}) by specializing the parameters as (\ref{para_odd}). Note
that the hypergeometric $\tau$ functions can be rewritten as
\begin{equation}
\begin{array}{ll}
\tau_n^{0,-1,-1}=\tau_n(a+1,b+1,c)\!\!\!
&=\tau_n(-r-N+2,-N+1,-r+1)\\[1mm]
&=\tau_n(-N+1,-r-N+2,-r+1)=\tau_n^{[1,2,1]},
\end{array}
\end{equation}
this time. In the case where $n'=m$, one can similarly verify the relation (\ref{f-v}) by using the
bilinear relations (\ref{bi_m}) and (\ref{bi_m'}).

Next, we prove that (\ref{eq-f}) is satisfied, which is rewritten by using (\ref{f-v}) as 
\begin{equation}
-r\dfrac{f_{n,m}}{v_{n,m}}
=\dfrac{nt^{-\frac{1}{2}}}{v_{n+1,m}^{-1}+v_{n-1,m}^{-1}}
+\dfrac{m}{v_{n,m+1}^{-1}+v_{n,m-1}^{-1}}.
\label{eq-f'}
\end{equation}
We use the bilinear relations 
\begin{equation}
\begin{array}{l}
n'\tau_{n'}^{0,0,0}\tau_{n'}^{0,-1,-1}
=(b-c+1)\tau_{n'+1}^{0,-1,0}\tau_{n'-1}^{0,0,-1}
+at^{-1}\tau_{n'+1}^{-1,-1,-1}\tau_{n'-1}^{1,0,0},\\[2mm]
(a+b-c+n'+1)\tau_{n'}^{0,0,0}\tau_{n'}^{0,-1,-1}
=a\tau_{n'}^{-1,-1,-1}\tau_{n'}^{1,0,0}
+(b-c+1)\tau_{n'}^{0,-1,0}\tau_{n'}^{0,0,-1},
\end{array}   \label{bi_BT'}
\end{equation}
and 
\begin{equation}
\begin{array}{l}
\tau_n^{0,0,0}\tau_n^{-1,-1,-2}
=-t^{-1}\tau_{n+1}^{-1,-1,-1}\tau_{n-1}^{0,0,-1}
+\tau_n^{-1,-1,-1}\tau_n^{0,0,-1},\\[2mm]
\tau_m^{0,0,0}\tau_m^{1,-1,0}
=\tau_m^{0,-1,0}\tau_m^{1,0,0}-\tau_{m+1}^{0,-1,0}\tau_{m-1}^{1,0,0},\\[2mm]
\tau_m^{0,-1,-1}\tau_m^{-1,0,-1}
=-\tau_{m+1}^{-1,-1,-1}\tau_{m-1}^{0,0,-1}+\tau_m^{-1,-1,-1}\tau_m^{0,0,-1},
\end{array}   \label{bi_constraint1}
\end{equation}
for the proof. Their derivation is also shown in Appendix \ref{sym_P6}. Let us consider the case
where $n'=n$. When $n+m$ is even, we have
\begin{equation}
\begin{array}{l}
-n\tau_n^{[1,2,2]}\tau_n^{[1,1,1]}
=N\tau_{n+1}^{[1,1,2]}\tau_{n-1}^{[1,2,1]}
+Nt^{-1}\tau_{n+1}^{[0,1,1]}\tau_{n-1}^{[2,2,2]},\\[1mm]
m\tau_n^{[1,2,2]}\tau_n^{[1,1,1]}
=N\tau_n^{[0,1,1]}\tau_n^{[2,2,2]}
+N\tau_n^{[1,1,2]}\tau_n^{[1,2,1]},
\end{array}   \label{bi_BT:ne}
\end{equation}
from the bilinear relations (\ref{bi_BT'}) by specializing the parameters $a,b$ and $c$ as given in
(\ref{para_even}). These lead us to
\begin{equation}
\begin{array}{l}
v_{n+1,m}^{-1}+v_{n-1,m}^{-1}
=c_1^{-1}t^{-r+\frac{n+1}{2}}\dfrac{n}{N}
\dfrac{\tau_n^{[1,2,2]}\tau_n^{[1,1,1]}}
      {\tau_{n+1}^{[0,1,1]}\tau_{n-1}^{[1,2,1]}},\\[5mm]
v_{n,m+1}^{-1}+v_{n,m-1}^{-1}
=-c_1^{-1}t^{-r+\frac{n}{2}}\dfrac{m}{N}
\dfrac{\tau_n^{[1,2,2]}\tau_n^{[1,1,1]}}
      {\tau_n^{[0,1,1]}\tau_n^{[1,2,1]}}.
\end{array}
\end{equation}
By using 
\begin{equation}
\tau_n^{[1,2,2]}\tau_n^{[0,1,0]}
=-t^{-1}\tau_{n+1}^{[0,1,1]}\tau_{n-1}^{[1,2,1]}
+\tau_n^{[0,1,1]}\tau_n^{[1,2,1]},
\end{equation}
which is obtained from the first relation in (\ref{bi_constraint1}), one can verify
(\ref{eq-f'}). When $n+m$ is odd, we have the bilinear relations
\begin{equation}
\begin{array}{l}
-n\tau_n^{[2,2,2]}\tau_n^{[1,2,1]}
=(-r+N)\tau_{n+1}^{[1,2,2]}\tau_{n-1}^{[2,2,1]}
+(r+N-1)t^{-1}\tau_{n+1}^{[1,1,1]}\tau_{n-1}^{[2,3,2]},\\[1mm]
m\tau_n^{[2,2,2]}\tau_n^{[1,2,1]}
=(r+N-1)\tau_n^{[1,1,1]}\tau_n^{[2,3,2]}
+(-r+N)\tau_n^{[1,2,2]}\tau_n^{[2,2,1]},
\end{array}   \label{bi_BT:no}
\end{equation}
from (\ref{bi_BT'}) with (\ref{para_odd}), and 
\begin{equation}
\tau_n^{[2,2,2]}\tau_n^{[1,1,0]}
=-t^{-1}\tau_{n+1}^{[1,1,1]}\tau_{n-1}^{[2,2,1]}
+\tau_n^{[1,1,1]}\tau_n^{[2,2,1]},
\end{equation}
from the first relation in (\ref{bi_constraint1}). These lead us to (\ref{eq-f'}). We next consider
the case where $n'=m$. When $n+m$ is even, we get the bilinear relations
\begin{equation}
\begin{array}{l}
-m\tau_m^{[1,2,2]}\tau_m^{[1,1,1]}
=N\tau_{m+1}^{[1,1,2]}\tau_{m-1}^{[1,2,1]}
+Nt^{-1}\tau_{m+1}^{[0,1,1]}\tau_{m-1}^{[2,2,2]},\\[1mm]
n\tau_m^{[1,2,2]}\tau_m^{[1,1,1]}
=N\tau_m^{[0,1,1]}\tau_m^{[2,2,2]}
+N\tau_m^{[1,1,2]}\tau_m^{[1,2,1]},
\end{array}   \label{bi_BT:me}
\end{equation}
and 
\begin{equation}
\tau_m^{[1,2,2]}\tau_m^{[2,1,2]}
=\tau_m^{[1,1,2]}\tau_m^{[2,2,2]}
-\tau_{m+1}^{[1,1,2]}\tau_{m-1}^{[2,2,2]},
\end{equation}
from (\ref{bi_BT'}) and the second relation in (\ref{bi_constraint1}), respectively. By using these
relations, one can show (\ref{eq-f'}) in a similar way to the case where $n'=n$. When $n+m$ is odd,
we use the bilinear relations
\begin{equation}
\begin{array}{l}
-m\tau_{m}^{[2,2,2]}\tau_{m}^{[1,2,1]}
=(-r+N)\tau_{m+1}^{[1,2,2]}\tau_{m-1}^{[2,2,1]}
+(r+N-1)t^{-1}\tau_{m+1}^{[1,1,1]}\tau_{m-1}^{[2,3,2]},\\[1mm]
n\tau_{m}^{[2,2,2]}\tau_{m}^{[1,2,1]}
=(r+N-1)\tau_{m}^{[1,1,1]}\tau_{m}^{[2,3,2]}
+(-r+N)\tau_{m}^{[1,2,2]}\tau_{m}^{[2,2,1]},
\end{array}   \label{bi_BT:mo}
\end{equation}
and 
\begin{equation}
\tau_{m}^{[1,2,1]}\tau_{m}^{[2,1,1]}
=-\tau_{m+1}^{[1,1,1]}\tau_{m-1}^{[2,2,1]}
+\tau_{m}^{[1,1,1]}\tau_{m}^{[2,2,1]},
\end{equation}
which are obtained from (\ref{bi_BT'}) and the third relation in (\ref{bi_constraint1}),
respectively, to show (\ref{eq-f'}).

We next give the verification of the compatibility condition (\ref{compatibilty}) by using the
bilinear relations
\begin{equation}
\begin{array}{l}
(c-a)\tau_{n'}^{0,-1,-1}\tau_{n'+1}^{-1,-1,0}
-b\tau_{n'}^{0,0,0}\tau_{n'+1}^{-1,-2,-1}
=(t-1)\tau_{n'}^{-1,-1,-1}\tau_{n'+1}^{0,-1,0},\\[2mm]
(c-a)t\tau_{n'}^{0,-1,-1}\tau_{n'+1}^{-1,-1,0}
-b\tau_{n'}^{0,0,0}\tau_{n'+1}^{-1,-2,-1}
=(t-1)\tau_{n'}^{0,-1,0}\tau_{n'+1}^{-1,-1,-1}.
\end{array}   \label{bi_constraint2}
\end{equation}
The derivation of these is discussed in Appendix \ref{sym_P6}. We first consider the case where
$n'=n$. When $n+m$ is even, we get
\begin{equation}
\begin{array}{l}
(-r+N+1)\tau_{n}^{[1,1,1]}\tau_{n+1}^{[0,1,2]}
+(r+N)\tau_{n}^{[1,2,2]}\tau_{n+1}^{[0,0,1]}
=(t-1)\tau_{n}^{[0,1,1]}\tau_{n+1}^{[1,1,2]},\\[1mm]
(-r+N+1)t\tau_{n}^{[1,1,1]}\tau_{n+1}^{[0,1,2]}
+(r+N)\tau_{n}^{[1,2,2]}\tau_{n+1}^{[0,0,1]}
=(t-1)\tau_{n}^{[1,1,2]}\tau_{n+1}^{[0,1,1]},
\end{array}
\end{equation}
from the bilinear relations (\ref{bi_constraint2}). Then we have 
\begin{equation}
\begin{array}{l}
t^{\frac{1}{2}}v_{n,m}+v_{n+1,m+1}
=t^{-\frac{n+1}{2}}(t-1)
\dfrac{(r)_N}{(-r+1)_{N+1}}
\dfrac{\tau_{n}^{[1,1,2]}\tau_{n+1}^{[0,1,1]}}
      {\tau_{n}^{[1,2,2]}\tau_{n+1}^{[0,1,2]}},\\[5mm]
v_{n,m}+t^{\frac{1}{2}}v_{n+1,m+1}
=t^{-\frac{n}{2}}(t-1)
\dfrac{(r)_N}{(-r+1)_{N+1}}
\dfrac{\tau_{n}^{[0,1,1]}\tau_{n+1}^{[1,1,2]}}
      {\tau_{n}^{[1,2,2]}\tau_{n+1}^{[0,1,2]}},
\end{array}
\end{equation}
from which we arrive at the compatibility condition (\ref{compatibilty}). When $n+m$ is odd, we have 
\begin{equation}
\begin{array}{l}
N\tau_{n}^{[1,2,1]}\tau_{n+1}^{[1,1,2]}
+N\tau_{n}^{[2,2,2]}\tau_{n+1}^{[0,1,1]}
=(t-1)\tau_{n}^{[1,1,1]}\tau_{n+1}^{[1,2,2]},\\[1mm]
Nt\tau_{n}^{[1,2,1]}\tau_{n+1}^{[1,1,2]}
+N\tau_{n}^{[2,2,2]}\tau_{n+1}^{[0,1,1]}
=(t-1)\tau_{n}^{[1,2,2]}\tau_{n+1}^{[1,1,1]},
\end{array}
\end{equation}
from (\ref{bi_constraint2}). Calculating $t^{\frac{1}{2}}v_{n,m}+v_{n+1,m+1}$ and
$v_{n,m}+t^{\frac{1}{2}}v_{n+1,m+1}$ by means of these relations, we see that we have
(\ref{compatibilty}). In the case where $n'=m$, one can verify the compatibility condition
(\ref{compatibilty}) in a similar manner.

Let us finally verify the similarity condition (\ref{similarity}), which can be written as 
\begin{equation}
\dfrac{n}{2}-\dfrac{1}{2}\chi_{n+m}-t\dfrac{d}{dt}\log v_{n,m}
=\dfrac{nv_{n+1,m}}{v_{n+1,m}+v_{n-1,m}}.
\label{similarity'}
\end{equation}
Here, we take the factor $\chi_{n+m}$ as $\chi_{n+m}=r\left[(-1)^{n+m}-1\right]$. The relevant
bilinear relations for the hypergeometric $\tau$ function are
\begin{equation}
\begin{array}{l}
(D+n)\tau_n^{0,0,0}\cdot\tau_n^{0,-1,-1}
=at^{-1}\tau_{n+1}^{-1,-1,-1}\tau_{n-1}^{1,0,0},\\[1mm]
(D+b-c+1)\tau_m^{0,-1,-1}\cdot\tau_m^{0,0,0}
=(b-c+1)\tau_m^{0,-1,0}\tau_m^{0,0,-1},\\[1mm]
(D+a+m)\tau_m^{0,0,0}\cdot\tau_m^{0,-1,-1}
=a\tau_m^{-1,-1,-1}\tau_m^{1,0,0}.
\end{array}   \label{bi_D'}
\end{equation}
The derivation of these is obtained in Appendix \ref{sym_P6}. We first consider the case where
$n'=n$. When $n+m$ is even, it is easy to see that we have
\begin{equation}
n\dfrac{v_{n+1,m}}{v_{n+1,m}+v_{n-1,m}}=-Nt^{-1}
\dfrac{\tau_{n+1}^{[0,1,1]}\tau_{n-1}^{[2,2,2]}}
      {\tau_{n}^{[1,2,2]}\tau_{n}^{[1,1,1]}},
\end{equation}
from the bilinear relation (\ref{bi_BT:ne}). We get 
\begin{equation}
(D+n)\tau_{n}^{[1,2,2]}\cdot\tau_{n}^{[1,1,1]}
=-Nt^{-1}\tau_{n+1}^{[0,1,1]}\tau_{n-1}^{[2,2,2]},
\end{equation}
from the first relation in (\ref{bi_D'}) with (\ref{para_even}). From this we can obtain the
similarity condition (\ref{similarity'}) as follows. When $n+m$ is odd, we have
\begin{equation}
(D+n)\tau_{n}^{[2,2,2]}\cdot\tau_{n}^{[1,2,1]}
=-t^{-1}(r+N-1)\tau_{n+1}^{[1,1,1]}\tau_{n-1}^{[2,3,2]},
\end{equation}
from the first relation in (\ref{bi_D'}). This relation together with the first relation in
(\ref{bi_BT:no}) leads us to (\ref{similarity'}). Next, we discuss the case where $n'=m$. When $n+m$
is even, we have
\begin{equation}
(D+N)\tau_{m}^{[1,2,2]}\cdot\tau_{m}^{[1,1,1]}
=N\tau_{m}^{[1,1,2]}\tau_{m}^{[1,2,1]},
\end{equation}
from the second relation in (\ref{bi_D'}). Then we arrive at (\ref{similarity'}) by virtue of the
second relation in (\ref{bi_BT:me}). When $n+m$ is odd, we get
\begin{equation}
\textstyle
(D+r+\frac{n-m-1}{2})
\tau_{m}^{[1,2,1]}\cdot\tau_{m}^{[2,2,2]}
=(r+N-1)\tau_{m}^{[1,1,1]}\tau_{m}^{[2,3,2]},
\end{equation}
from the third relation in (\ref{bi_D'}). Then we derive the similarity condition
(\ref{similarity'}) by using the second relation in (\ref{bi_BT:mo}). This completes the proof of
Theorem \ref{formula_f} and Proposition \ref{formula_v}.

\section{Extension of the domain}
First, we extend the domain of the discrete power function to $\mathbb{Z}^2$. To determine the
values of $f_{n,m}$ in the second, third and fourth quadrants, we have to give the values of
$f_{-1,0}$ and $f_{0,-1}$ as the initial conditions. Set the initial conditions as
\begin{equation}
f_{-1,0}=c_2t^{2r},\quad f_{0,-1}=c_3t^{3r},\label{init_ext}
\end{equation}
where $c_2$ and $c_3$ are arbitrary constants. This is natural because these conditions reduce to 
\begin{equation}
f_{1,0}=1,\quad f_{0,1}=e^{\pi ir},\quad
f_{-1,0}=e^{2\pi ir},\quad f_{0,-1}=e^{3\pi ir}
\end{equation}
at the original setting. Due to the symmetry of equations (\ref{LS-KdV}) and (\ref{eq-f}), we
immediately obtain the explicit formula of $f_{n,m}$ in the second and third quadrant.

\begin{corollary}
Under the initial conditions $f_{0,1}=c_1t^r$ and (\ref{init_ext}), we have 
\begin{equation}
\begin{array}{l}
f_{-n,m}=f_{n,m}|_{\,c_0\mapsto c_2t^{2r}},\quad
f_{-n,-m}=f_{n,m}|_{\,c_0\mapsto c_2t^{2r},c_1\mapsto c_3t^{2r}},
\end{array}
\end{equation}
for $n,m\in\mathbb{Z}_+$. 
\end{corollary}

\begin{figure}
\centering
\begin{tabular}{cc}
\begin{minipage}{0.4\textwidth}
\vskip-10mm
\begin{center}
\includegraphics[scale=0.2]{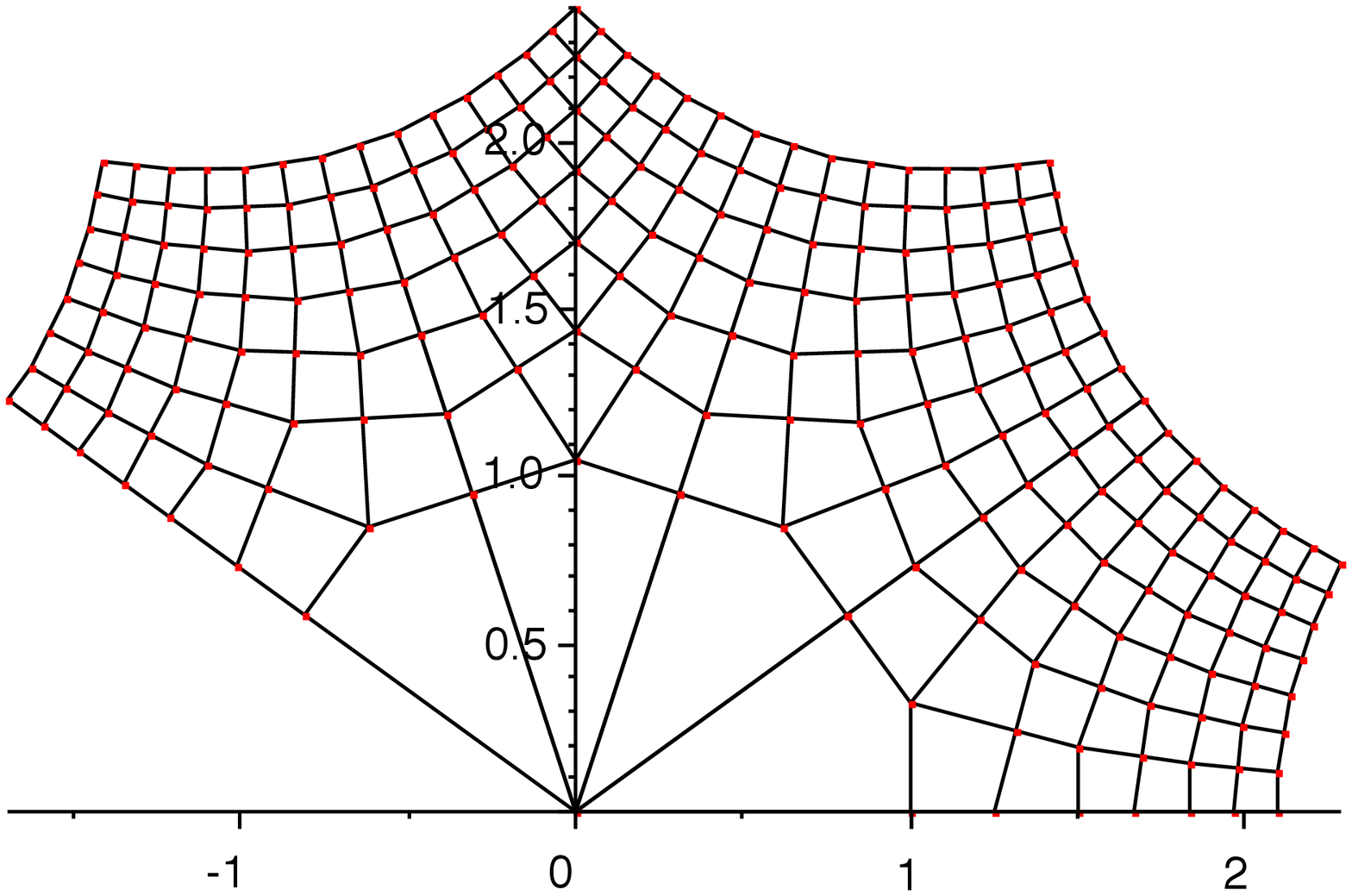}
\caption{\small{The discrete power function with $\gamma=5/2$ whose domain is $\mathbb{Z}^2$.}}
\label{Z2}
\end{center}
\end{minipage}
&
\begin{minipage}{0.4\textwidth}
\vskip9.5mm
\begin{center}
\includegraphics[scale=0.23]{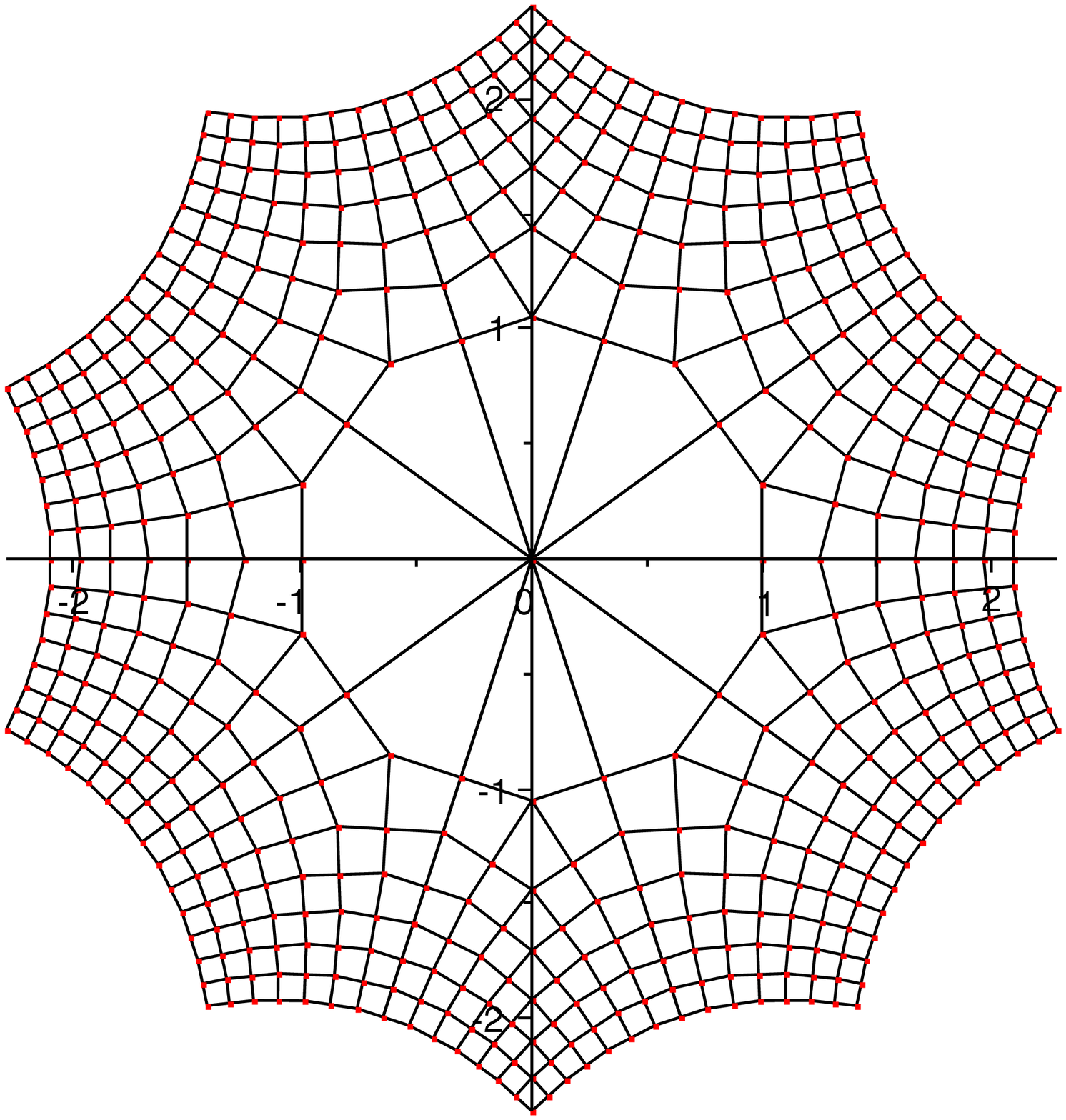}
\caption{\small{The discrete power function with $\gamma=5/2$ whose domain is the discrete Riemann surface.}}
\label{Riemann}
\end{center}
\end{minipage}
\end{tabular}
\end{figure}

Next, let us discuss the explicit formula in the fourth quadrant. Naively, we use the initial
conditions $f_{0,-1}=c_3t^{3r}$ and $f_{1,0}=c_0$ to get the formula
$f_{n,-m}=f_{n,m}|_{\,c_1\mapsto c_3t^{2r}}$. However, this setting makes the discrete power
function $f_{n,m}$ become a single-valued function on $\mathbb{Z}^2$. In order to allow $f_{n,m}$ to
be multi-valued on $\mathbb{Z}^2$, we introduce a discrete analogue of the Riemann surface by the
following procedure. Prepare an infinite number of $\mathbb{Z}^2$-planes, cut the positive part of
the ``real axis'' of each $\mathbb{Z}^2$-plane and glue them in a similar way to the continuous
case. The next step is to write the initial conditions (\ref{initial}) and (\ref{init_ext}) in polar
form as
\begin{equation}
f(1,\pi k/2)=c_kt^{kr}\quad(k=0,1,2,3),
\end{equation}
where the first component, $1$, denotes the absolute value of $n+im$ and the second component, $\pi
k/2$, is the argument. We must generalize the above initial conditions to those for arbitrary
$k\in\mathbb{Z}$ so that we obtain the explicit expression of $f_{n,m}$ for each quadrant of each
$\mathbb{Z}^2$-plane. Let us illustrate a typical case. When $\dfrac{3}{2}\pi\le\arg(n+im)\le 2\pi$,
we solve the equations (\ref{LS-KdV}) and (\ref{eq-f}) under the initial conditions
\begin{equation}
f(1,3\pi/2)=c_3t^{3r},\quad f(1,2\pi)=c_4t^{4r},
\end{equation}
to obtain the formula
\begin{equation}
f_{-n,-m}=f_{n,m}|_{\,c_0\mapsto c_4t^{4r},c_1\mapsto c_3t^{2r}}
\quad(n,m\in\mathbb{Z}_+).
\end{equation}
We present the discrete power function with $\gamma=5/2$ whose domain is $\mathbb{Z}^2$ and the
discrete Riemann surface in Figure \ref{Z2} and \ref{Riemann}, respectively. Note that the necessary
and sufficient condition for the discrete power function to reduce to a single-valued function on
$\mathbb{Z}^2$ is ($c_k=c_{k+4}$ and) $e^{4\pi ir}=1$, which means that the exponent $\gamma$ is an
integer.

\section{Associated circle pattern of Schramm type}
Agafonov and Bobenko have shown that the discrete power function for real $\gamma$ is an immersion
and thus defines a circle pattern of Schramm type. We have generalized the discrete power function
to complex $\gamma$. It is natural to ask whether there are other cases where the discrete power
function is associated with circle patterns. We have the following result:
\begin{theorem}\label{thm:immersion}
The mapping $f:\, \mathbb{Z}^2_+\to\mathbb{C}\,;(n,m)\mapsto f_{n,m}$ satisfying
(\ref{cr}) and (\ref{eq-f}) with the initial condition
\begin{equation}\label{initial_f_immsesion}
 f_{0,0}=0,\quad f_{1,0}=1,\quad f_{0,1}=e^{\pi i \gamma/2},\quad \gamma\in\mathbb{C}
\end{equation}
is an immersion when ${\rm Re}~\gamma=1$. 
\end{theorem}
In this section, we give the proof of Theorem \ref{thm:immersion} along with the discussion in
\cite{Agafonov-Bobenko:2000}. We also use the explicit formulae given in previous sections.
\newcommand{\BR}{\mathbb{R}}
\subsection{Circle pattern}
Setting
\begin{equation}\label{Re_gamma=1}
 \gamma = 1 + i\delta,\quad \delta\in\mathbb{R},
\end{equation}
we associate the discrete power function with circle patterns of Schramm type. The proof of Theorem \ref{thm:immersion} is then reduced to properties of the radii of those circles.
\begin{lem}\label{lem:equidistance}
A discrete power function $f_{n,m}$ defined by (\ref{cr}) and (\ref{eq-f}) with initial condition
\begin{equation}\label{initial_f}
f_{0,0}=0,\quad f_{1,0}=1,\quad f_{0,1}=c_1e^{\pi i\gamma/2},\quad c_1>0
\end{equation}
for arbitrary $\gamma\in\mathbb{C}\backslash{2\mathbb{Z}}$ has the
equidistant property
\begin{equation}\label{equi_f:1}
f_{2n,0}-f_{2n-1,0}=f_{2n+1,0}-f_{2n,0},\quad
f_{0,2m}-f_{0,2m-1}=f_{0,2m+1}-f_{0,2m}, 
\end{equation}
for any $n\ge1,m\ge1$. Moreover, if and only if ${\rm Re}~\gamma=1$, we have
\begin{equation}\label{equi_f:2}
|f_{n+1,0}-f_{n,0}|=|f_{n,0}-f_{n-1,0}|,\quad
|f_{0,m+1}-f_{0,m}|=|f_{0,m}-f_{0,m-1}|, 
\end{equation}
for any $n\ge1,m\ge1$. 
\end{lem}
\noindent\textbf{Proof.}\quad By using the formulae in Theorem \ref{formula_f} (or Proposition
\ref{formula_v}), we have
\begin{equation}
\begin{split}
& f_{2n,0}-f_{2n-1,0} =
  f_{2n+1,0}-f_{2n,0}=\frac{\gamma}{2n+\gamma}\prod_{k=1}^n\frac{2k+\gamma}{2k-\gamma},\\ 
& f_{0,2m}-f_{0,2m-1} =
  f_{0,2m+1}-f_{0,2m}=c_1e^{\frac{\pi i\gamma}{2}}\frac{\gamma}{2m+\gamma}\prod_{k=1}^m\frac{2k+\gamma}{2k-\gamma},
\end{split}
\end{equation}
which proves (\ref{equi_f:1}). We also have
\begin{equation}\label{equi_f:3}
f_{2n+2,0}-f_{2n+1,0}=\prod_{k=1}^{n+1}\frac{2k-2+\gamma}{2k-\gamma},
\quad
 f_{2n+1,0}-f_{2n,0}=\prod_{k=1}^{n}\frac{2k-2+\gamma}{2k-\gamma}.
\end{equation}
Putting $\gamma=1+i\delta$, we obtain
\begin{equation}
 f_{2n+2,0}-f_{2n+1,0}=\prod_{k=1}^{n+1}\frac{2k-1+i\delta}{2k-1-i\delta},\quad
 f_{2n+1,0}-f_{2n,0}=\prod_{k=1}^{n}\frac{2k-1+i\delta}{2k-1-i\delta},
\end{equation}
which implies $|f_{2n+2,0}-f_{2n+1,0}|=|f_{2n+1,0}-f_{2n,0}|=1$.
Using the first equation of (\ref{equi_f:1}), we see that the first equation of
(\ref{equi_f:2}) follows. The second equation of (\ref{equi_f:2}) can be proved in a similar
manner. Suppose that (\ref{equi_f:2}) holds, then from (\ref{equi_f:3}) we have
\begin{equation}
 \left|\frac{f_{2n+2,0}-f_{2n+1,0}}{f_{2n+1,0}-f_{2n,0}}\right|
=\left|\frac{2n+\gamma}{2n+2-\gamma}\right|=1,
\end{equation}
which leads us to ${\rm Re}~\gamma=1$.\qed
\begin{prop}\label{prop:kite}
Let $f_{n,m}$ satisfy (\ref{cr}) and (\ref{eq-f}) in $\BZ^2_+$ with initial condition
(\ref{initial_f}). Then all the elementary quadrilaterals
$(f_{n,m},f_{n+1,m},f_{n+1,m+1},f_{n,m+1})$ are of the kite form, namely, all edges at each vertex
$f_{n,m}$ with $n+m=1\,(\mbox{mod}\,2)$ are of the same length,
\begin{equation}
|f_{n+1,m}-f_{n,m}|=|f_{n,m+1}-f_{n,m}|=
|f_{n-1,m}-f_{n,m}|=|f_{n,m-1}-f_{n,m}|. 
\end{equation}
Moreover, all angles between the neighbouring edges at the vertex $f_{n,m}$ with
$n+m=0\,(\mbox{mod}\,2)$ are equal to $\pi/2$.
\end{prop}
\noindent\textbf{Proof.}\quad For three complex numbers $z_i$ ($i=1,2,3$),
we introduce a notation
\begin{equation}
 [z_1,z_2,z_3] = \arg \frac{z_1-z_2}{z_3-z_2}.
\end{equation}
We first consider the quadrilateral $(f_{0,0},f_{1,0},f_{1,1},f_{0,1})$. Notice that $f_{0,1}=ic_1e^{-\frac{\pi\delta}{2}}$ which implies that $[f_{0,1},f_{0,0},f_{1,0}]=\frac{\pi}{2}$. Then it follows from (\ref{cr}) that $[f_{1,0},f_{1,1},f_{0,1}]=\frac{\pi}{2}$, $|f_{1,0}-f_{0,0}|=|f_{1,1}-f_{1,0}|$ and $|f_{0,1}-f_{0,0}|=|f_{1,1}-f_{0,1}|$. We next consider the quadrilateral $(f_{1,0},f_{2,0},f_{2,1},f_{1,1})$ where, from Lemma \ref{lem:equidistance}, we have $|f_{1,1}-f_{1,0}|=|f_{2,0}-f_{1,0}|$. We see from (\ref{cr}) that $[f_{2,1},f_{1,1},f_{1,0}]=[f_{1,0},f_{2,0},f_{2,1}]=\frac{\pi}{2}$ and $|f_{2,1}-f_{1,1}|=|f_{2,1}-f_{2,0}|$. From Lemma \ref{lem:equidistance} and $[f_{1,0},f_{2,0},f_{2,1}]=\frac{\pi}{2}$, we see that $[f_{2,1},f_{2,0},f_{3,0}]=\frac{\pi}{2}$. Then a similar argument can be applied to the quadrilateral $(f_{2,0},f_{3,0},f_{3,1},f_{2,1})$ and so forth. In this manner, Proposition \ref{prop:kite} is proved inductively.\qed
\par\bigskip

\begin{figure}
\centering
\setlength{\unitlength}{1cm}
\includegraphics[scale=0.3]{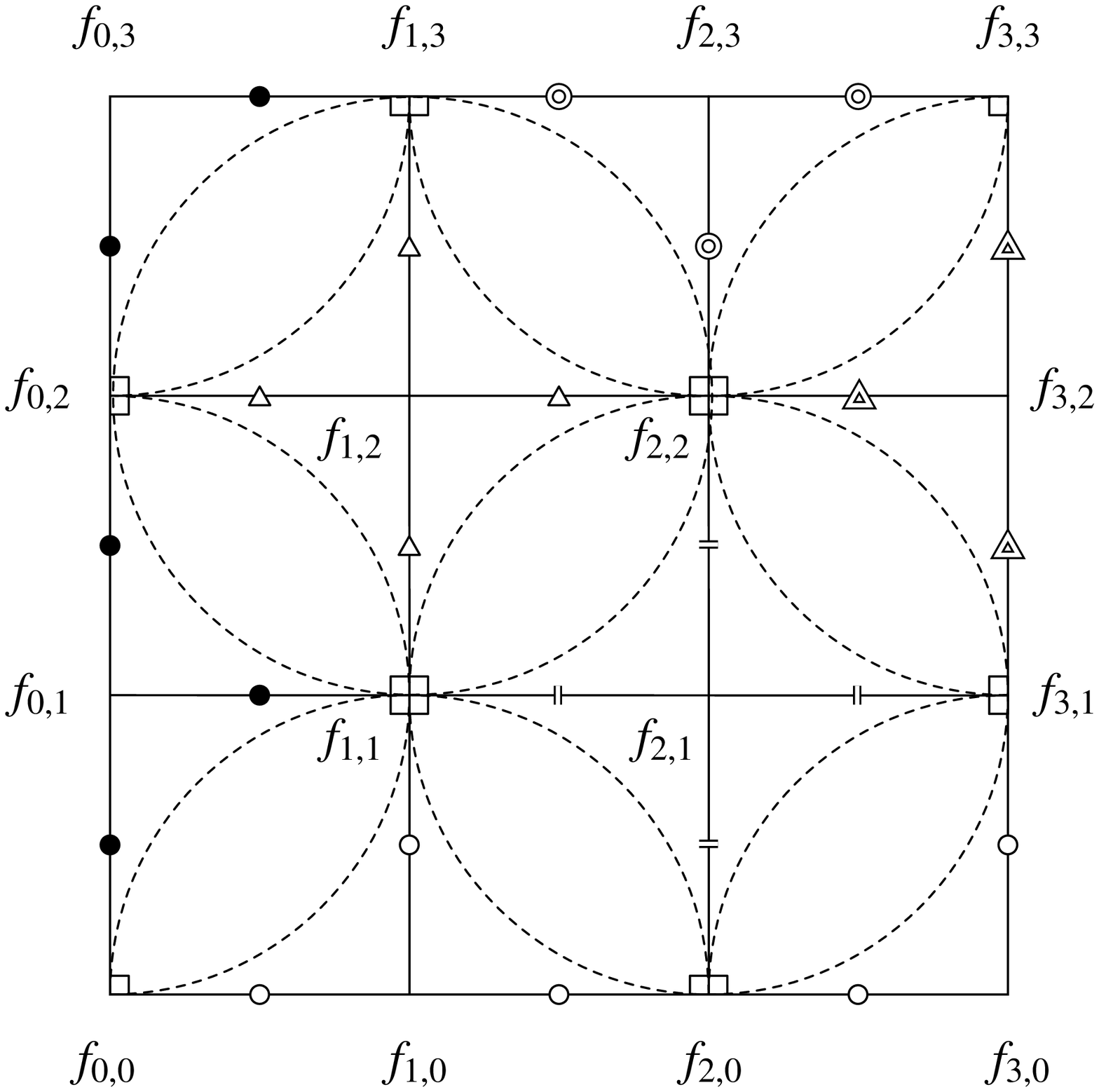}
\caption{{\small A schematic diagram of quadrilaterals.}}
\label{fig:f_config_init}
\end{figure}

From Proposition \ref{prop:kite}, it follows that the circumscribed circles of the quadrilaterals
\hfill\break $(f_{n-1,m},f_{n,m-1},f_{n+1,m},f_{n,m+1})$ with $n+m=1\,(\mbox{mod}\,2)$ form a circle
pattern of Schramm type \cite{Agafonov-Bobenko:2000,Schramm}, namely, the circles of neighbouring
quadrilaterals intersect orthogonally and the circles of half-neighbouring quadrilaterals with a common vertex are tangent (See Figure \ref{fig:f_config_init}).
Conversely, for a given circle pattern of Schramm type, it is possible to construct a discrete
conformal mapping $f_{n,m}$ as follows.  Let $\{C_{n,m}\},\,(n,m)\in{\bf
V}=\{(n,m)\in\BZ_+^2\,|\,n+m=1\,(\mbox{mod}\,2)\}$ be a circle pattern of Schramm type on the
complex plane. Define $f\,:\,\BZ_+^2\to\BC\,;\,(n,m)\mapsto f_{n,m}$ in the following manner:
\begin{itemize}
\item[(a)] If $n+m=1\,(\mbox{mod}\,2)$, then $f_{n,m}$ is the center of $C_{n,m}$. 
\item[(b)] If $n+m=0\,(\mbox{mod}\,2)$, then $f_{n,m}:=C_{n-1,m}\cap C_{n+1,m}=C_{n,m+1}\cap C_{n,m-1}$. 
\end{itemize}
By construction, it follows that all elementary quadrilaterals
$(f_{n,m},f_{n+1,m},f_{n+1,m+1},f_{n,m+1})$ are of the kite form whose angles between the edges
with different lengths are $\frac{\pi}{2}$. Therefore, (\ref{cr}) is satisfied automatically. In what
follows, the function $f_{n,m}$, defined by (a) and (b), is called a discrete conformal map
corresponding to the circle pattern $\{C_{n,m}\}$.

We now use the the radii of corresponding circle patterns to characterize the necessary and sufficient condition that the discrete power function is an immersion.

\begin{thm}\label{thm:immersion_R}
Let $f_{n,m}$ satisfying (\ref{cr}) and (\ref{eq-f}) with initial condition
(\ref{initial_f}) be an immersion. Then $R_{n,m}$ defined by
\begin{equation}
R_{n,m}=|f_{n+1,m}-f_{n,m}|=|f_{n,m+1}-f_{n,m}|
=|f_{n-1,m}-f_{n,m}|=|f_{n,m-1}-f_{n,m}| 
\end{equation}
satisfies
\begin{equation}
n\dfrac{R_{n+1,m}-R_{n-1,m}}{R_{n+1,m}+R_{n-1,m}}+
m\dfrac{R_{n,m+1}-R_{n,m-1}}{R_{n,m+1}+R_{n,m-1}}=0,\label{eq_1}
\end{equation}
and
\begin{equation}
R_{n+1,m+2}=
\dfrac{[(m+1)R_{n,m+1}+\delta R_{n+1,m}]R_{n,m+1}(R_{n+1,m}+R_{n-1,m})
       +nR_{n+1,m}(R_{n,m+1}^2-R_{n+1,m}R_{n-1,m})}
      {[(m+1)R_{n+1,m}-\delta R_{n,m+1}](R_{n+1,m}+R_{n-1,m})
       -n(R_{n,m+1}^2-R_{n+1,m}R_{n-1,m})},
\label{eq_2}
\end{equation}
for $(n,m)\in{\bf V}$. Conversely, let $R\,:\,{\bf V}\to\BR_+$ satisfy (\ref{eq_1}) and
 (\ref{eq_2}). Then $R_{n,m}$ defines an immersed circle pattern of Schramm type. The corresponding
 discrete conformal map $f_{n,m}$ is an immersion and satisfies (\ref{eq-f}).
\end{thm}

\begin{figure}
\centering
\includegraphics[scale=0.3]{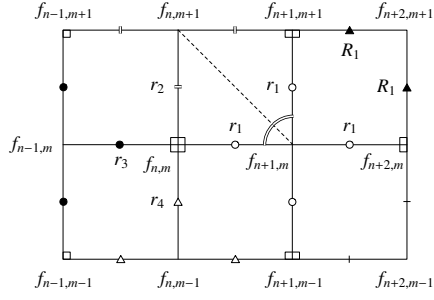}
\caption{{\small Parametrization of edges.}}
\label{fig:f_config_fr}
\end{figure}

\noindent\textbf{Proof.}  The proof of Theorem \ref{thm:immersion_R} occupies the remainder of this subsection. Suppose that the discrete power function $f_{n,m}$ is
immersed. For $n+m=0\,(\mbox{mod}\,2)$, we may parametrize the edges around the vertex $f_{n,m}$ as
\begin{equation}\label{eqn:edges_param1}
f_{n+1,m}-f_{n,m}=r_1e^{i\beta},\quad f_{n,m+1}-f_{n,m}=ir_2e^{i\beta},\quad 
f_{n-1,m}-f_{n,m}=-r_3e^{i\beta},\quad f_{n,m-1}-f_{n,m}=-ir_4e^{i\beta}, 
\end{equation}
where $r_i>0$ ($i=1,2,3,4$) are the radii of the corresponding circles,
since all the angles around $f_{n,m}$ are $\frac{\pi}{2}$. The constraint (\ref{eq-f}) reads 
\begin{equation}\label{eqn:edges_param2}
\gamma f_{n,m}=e^{i\beta}
\left(2n\dfrac{r_1r_3}{r_1+r_3}+2im\dfrac{r_2r_4}{r_2+r_4}\right). 
\end{equation}
\begin{lem}\label{lem:kite}
For $n+m=0\,(\mbox{mod}\,2)$ we have:
\begin{equation}\label{eqn:kite1}
f_{n+1,m+1}-f_{n+1,m}=-e^{i\beta}r_1\dfrac{r_1-ir_2}{r_1+ir_2},\quad
f_{n+1,m}-f_{n+1,m-1}=e^{i\beta}r_1\dfrac{r_1+ir_4}{r_1-ir_4}. 
\end{equation} 
\end{lem}
\noindent\textbf{Proof.} The kite form of the quadrilateral
$(f_{n,m},f_{n+1,m},f_{n+1,m+1},f_{n,m+1})$ implies
$f_{n+1,m+1}-f_{n+1,m}=-(f_{n+1,m}-f_{n,m})e^{-2i[f_{n,m+1},f_{n+1,m},f_{n+1,m+1}]}$
(See Figure \ref{fig:f_config_fr}). 
The first equation of (\ref{eqn:kite1}) follows by noticing that
$\tan[f_{n,m+1},f_{n+1,m},f_{n+1,m+1}]=\frac{r_2}{r_1}$. The second equation is derived by a similar
consideration on the quadrilateral $(f_{n,m-1},f_{n+1,m-1},f_{n+1,m},f_{n,m})$.\qed
\par\bigskip

\noindent Setting $f_{n+2,m}-f_{n+1,m}=r_1e^{i(\beta+\sigma)}$ and substituting
(\ref{eqn:edges_param1})--(\ref{eqn:kite1}) into (\ref{eq-f}) at the point $(n+1,m)$, one arrives at
\begin{equation}
2n\dfrac{r_1r_3}{r_1+r_3}+2im\dfrac{r_2r_4}{r_2+r_4}+\gamma r_1
=2(n+1)r_1\dfrac{1}{1+e^{-i\sigma}}
+2mr_1\dfrac{r_1(r_2-r_4)+i(r_1^2+r_2r_4)}{2r_1(r_2+r_4)}.
\label{eq_r}
\end{equation}
The real part of (\ref{eq_r}) gives
\begin{equation}
n\dfrac{r_1-r_3}{r_1+r_3}+m\dfrac{r_2-r_4}{r_2+r_4}=0, 
\end{equation}
which coincides with (\ref{eq_1}). 

\medskip

\noindent Now we parametrize the edges around the vertex $f_{n+1,m+1}$ with $n+m=0~({\rm mod}~2)$ as
\begin{equation}
\begin{split}
&f_{n+2,m+1}=f_{n+1,m+1}+R_1e^{i\beta'},\quad
f_{n+1,m+2}=f_{n+1,m+1}+iR_2e^{i\beta'},\\[2mm]
&f_{n,m+1}=f_{n+1,m+1}-r_2e^{i\beta'},\quad
f_{n+1,m}=f_{n+1,m+1}-ir_1e^{i\beta'}.
\end{split} 
\end{equation}
From the first equation of (\ref{eqn:kite1}) and noticing that all angles around the vertex
$f_{n+1,m+1}$ are $\frac{\pi}{2}$, we have the following relation between $\beta'$ and $\beta$:
\begin{equation}
e^{i\beta'}=ie^{i\beta}\dfrac{r_1-ir_2}{r_1+ir_2}. 
\end{equation}
One can express $f_{n+2,m+1}$ in two ways as (See Figure \ref{fig:f_config_fr})
\begin{equation}
f_{n+2,m+1}=f_{n+2,m}+iR_1e^{i(\beta+\sigma)}
=f_{n+1,m}+r_1e^{i(\beta+\sigma)}+iR_1e^{i(\beta+\sigma)},
\end{equation}
and
\begin{equation}
f_{n+2,m+1}=f_{n+1,m+1}+R_1e^{i\beta'}
=f_{n+1,m}-e^{i\beta}r_1\dfrac{r_1-ir_2}{r_1+ir_2}+R_1e^{i\beta'}. 
\end{equation}
The compatibility implies
\begin{equation}
e^{i\sigma}=\dfrac{R_1+ir_1}{R_1-ir_1}\dfrac{r_1-ir_2}{r_1+ir_2}. 
\end{equation}
Then, from the imaginary part of (\ref{eq_r}), one obtains
\begin{equation}\label{eqn:nmr}
m\dfrac{r_2r_4-r_1^2}{r_2+r_4}+\delta r_1
=(n+1)\dfrac{r_1^2-r_2R_1}{r_2+R_1}, 
\end{equation}
or solving (\ref{eqn:nmr}) with respect to $R_1=R_{n+2,m+1}$
\begin{equation}
R_{n+2,m+1}=
\dfrac{[(n+1)R_{n+1,m}-\delta R_{n,m+1}]R_{n+1,m}(R_{n,m+1}+R_{n,m-1})
       +mR_{n,m+1}(R_{n+1,m}^2-R_{n,m+1}R_{n,m-1})}
      {[(n+1)R_{n,m+1}+\delta R_{n+1,m}](R_{n,m+1}+R_{n,m-1})
       -m(R_{n+1,m}^2-R_{n,m+1}R_{n,m-1})}.
\label{(2,1)}
\end{equation}
We may rewrite (\ref{eq-f}) at $(n+1,m+1)$ in terms of $r_i$ ($i=1,2,3,4$) $R_1$, $R_2$ as
\begin{equation}\label{eqn:nmr2}
\begin{split}
&2n\dfrac{r_1r_3}{r_1+r_3}+2im\dfrac{r_2r_4}{r_2+r_4}
+\gamma r_1\left(1-\dfrac{r_1-ir_2}{r_1+ir_2}\right)\\[4mm]
&=i\dfrac{r_1-ir_2}{r_1+ir_2}\left[
2(n+1)\dfrac{r_2R_1}{r_2+R_1}+2i(m+1)\dfrac{r_1R_2}{r_1+R_2}\right].
\end{split}
\end{equation}
Eliminating $\gamma$ from (\ref{eq_r}) and (\ref{eqn:nmr2}), we get 
\begin{equation}
m\dfrac{r_2}{r_2+r_4}-n\dfrac{r_3}{r_1+r_3}
-(m+1)\dfrac{R_2}{r_1+R_2}+(n+1)\dfrac{r_2}{r_2+R_1}=0. 
\end{equation}
We then eliminate $R_1$ using (\ref{eqn:nmr}) to obtain
\begin{equation}
R_2=\dfrac{[(m+1)r_2+\delta r_1]r_2(r_1+r_3)+nr_1(r_2^2-r_1r_3)}
          {[(m+1)r_1-\delta r_2](r_1+r_3)-n(r_2^2-r_1r_3)}, 
\end{equation}
which coincides with (\ref{eq_2}). This proves the first part of Theorem \ref{thm:immersion_R}.
\medskip

\noindent To prove the second part, we use the following lemma. 
\begin{lem}
Let $R\,:\,\bf{V}\to\BR_+$ satisfy (\ref{eq_1}) and (\ref{eq_2}). 
Then, $R$ satisfies (\ref{(2,1)}) and 
\begin{equation}
\begin{array}{l}
\left[
(n+1)(R_{n+1,m}^2-R_{n,m-1}R_{n+2,m-1})
+\delta R_{n+1,m}(R_{n,m-1}+R_{n+2,m-1})\right](R_{n,m+1}+R_{n,m-1})\\[2mm]
-m(R_{n+1,m}^2-R_{n,m+1}R_{n,m-1})(R_{n,m-1}+R_{n+2,m-1})=0.
\end{array}   \label{(2,-1)}
\end{equation}
\end{lem}

\noindent \textbf{Proof.}\quad Substituting (\ref{eq_1}) at $(n,m)$ and
at $(n+1,m+1)$ into (\ref{eq_2}) to eliminate $R_{n-1,m}$ and
$R_{n+1,m+2}$, we get (\ref{(2,1)}). Substituting (\ref{eq_2}) at
$(n+1,m-1)$ into (\ref{(2,1)}), we get (\ref{(2,-1)}) under the condition $\delta
R_{n+1,m}(R_{n+2,m-1}-R_{n,m+1})+(n+m+1)(R_{n+2,m-1}R_{n,m+1}+R_{n+1,m}^2)\ne0$
which can be verified from the compatibility with (\ref{eq_1}) and (\ref{eq_2}).
\qed
\medskip

\noindent
Eliminating $\delta$ from (\ref{(2,1)}) and (\ref{(2,-1)}), we get 
\begin{equation}\label{5pts}
(R_{n+1,m}^2-R_{n,m+1}R_{n+2,m+1})(R_{n,m-1}+R_{n+2,m-1})
+(R_{n+1,m}^2-R_{n,m-1}R_{n+2,m-1})(R_{n,m+1}+R_{n+2,m+1})=0.
\end{equation}
In \cite{Schramm}, it was proven that, given $R_{n,m}$ satisfying (\ref{5pts}), the circle pattern
with radii of the circles $R_{n,m}$ is immersed. Thus, the corresponding discrete conformal map
$f_{n,m}$ is an immersion.

\begin{figure}
\centering
\begin{minipage}{0.4\textwidth}
\centering
\includegraphics[width=0.75\textwidth]{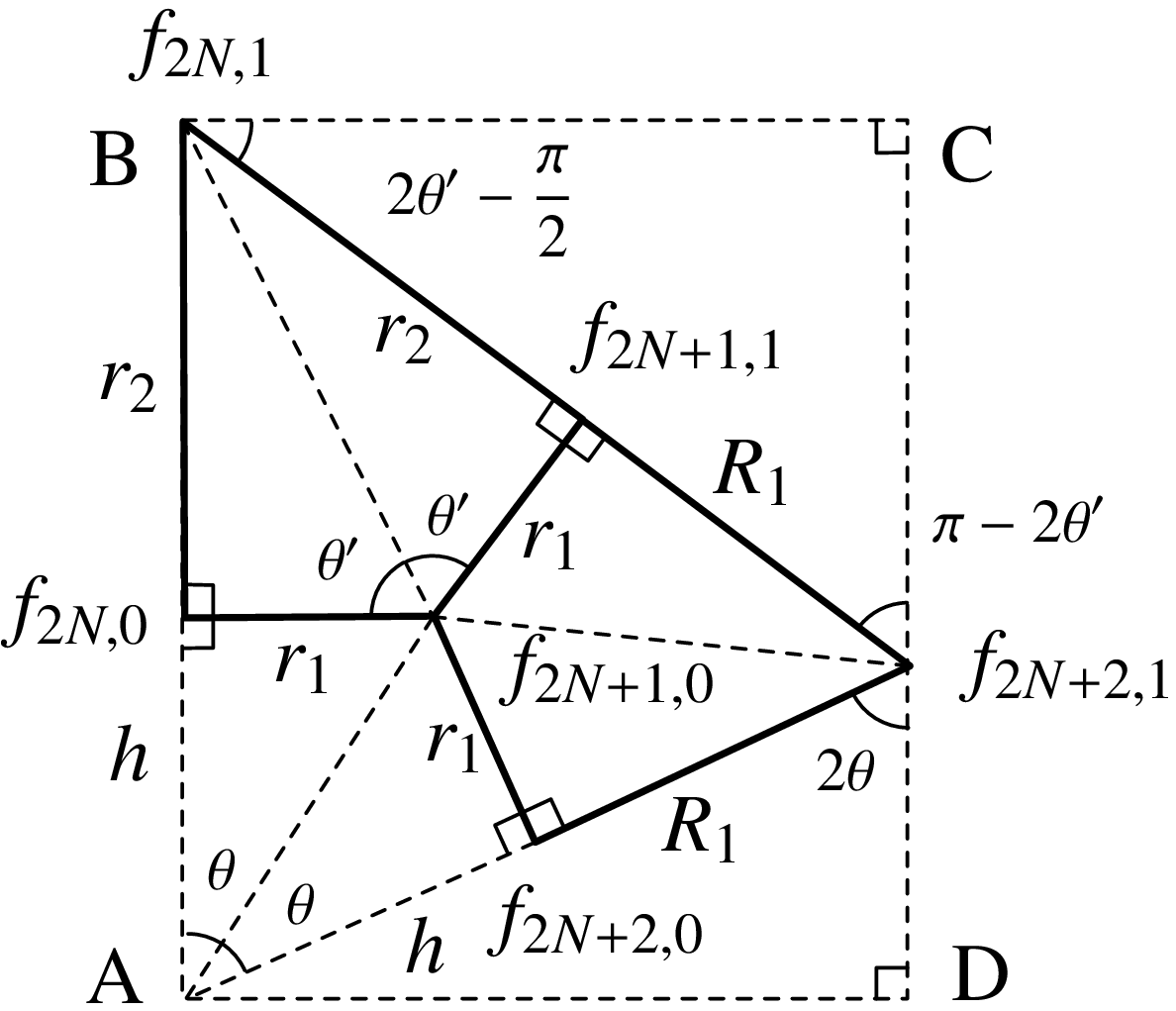}\\
(i) $-\frac{\pi}{2}\le2\theta\le0$
\end{minipage}
\quad
\begin{minipage}{0.4\textwidth}
\centering
\includegraphics[width=0.7\textwidth]{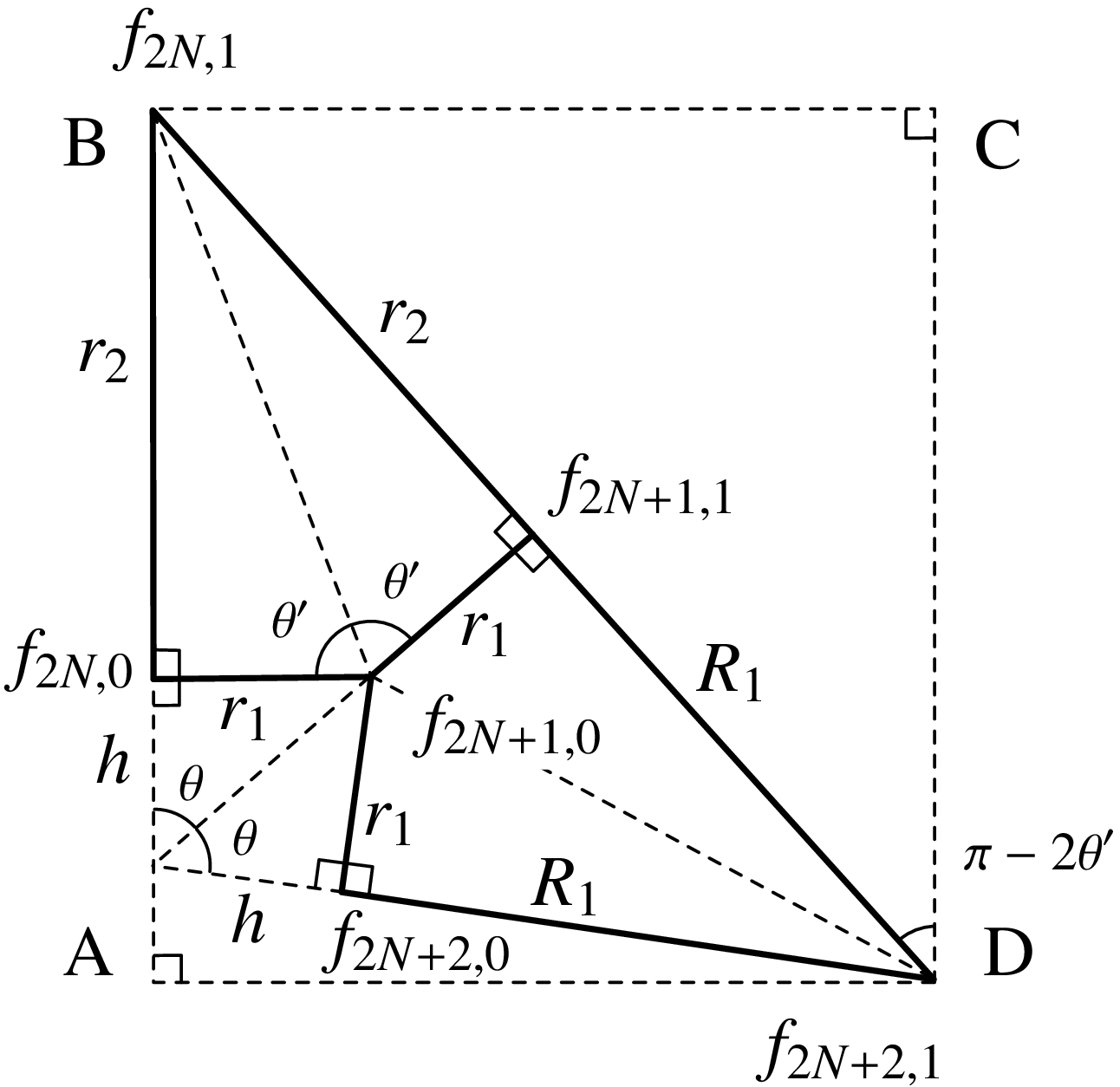}\\
(ii) $-\pi<2\theta\le-\frac{\pi}{2}$
\end{minipage}
\caption{{\small Configuration of points in Lemma \ref{angles}.}}\label{fig:delta_theta2}
\end{figure}

Let us finally show that the discrete conformal map $f_{n,m}$ satisfies (\ref{eq-f}). 
Putting $m=0$ in (\ref{eq_1}), we have $R_{n+1,0}=R_{n-1,0}$. This means that
$|f_{n+1,0}-f_{n,0}|=|f_{n,0}-f_{n-1,0}|$ for any $n\ge1$. By using the ambiguity of translation and
scaling of the circle pattern, one can set $f_{0,0}=0,\,f_{1,0}=1$ without loss of generality, and set
\begin{equation}\label{f_and_theta}
f_{2N+1,0}-f_{2N,0}=f_{2N,0}-f_{2N-1,0}=\exp\left(2i\sum_{j=1}^N\theta_j\right)
\quad(N=1,2,\ldots). 
\end{equation}
Putting $(n,m)=(2N,0)$ in (\ref{(2,1)}) we have
\begin{equation}\label{delta_R}
\delta=(2N+1)\dfrac{R_{2N+1,0}^2-R_{2N,1}R_{2N+2,1}}{R_{2N+1,0}(R_{2N,1}+R_{2N+2,1})}. 
\end{equation}
A geometric consideration leads us the following lemma:
\begin{lem}\label{angles}
We have
\begin{equation} 
\delta=(2N+1)\tan\theta_{N+1}\quad(N=0,1,\ldots).\label{theta}
\end{equation} 
\end{lem}
\noindent\textbf{Proof.}\quad 
\begin{figure}
\centering
\begin{minipage}{0.4\textwidth}
\centering
\includegraphics[width=0.95\textwidth]{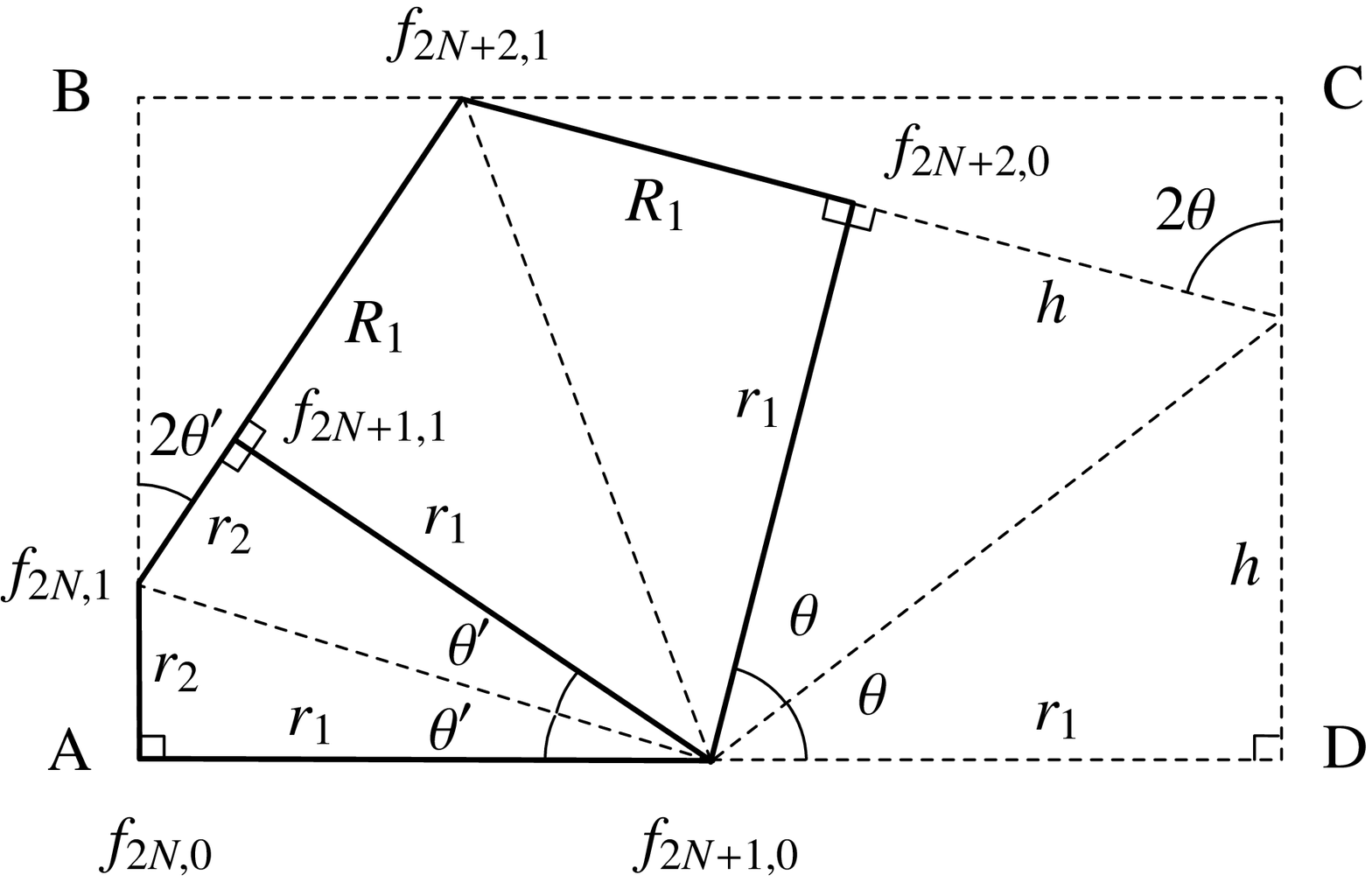}\\
(i) $0\le2\theta\le\frac{\pi}{2}$
\end{minipage}
\quad
\begin{minipage}{0.4\textwidth}
\centering
\includegraphics[width=0.95\textwidth]{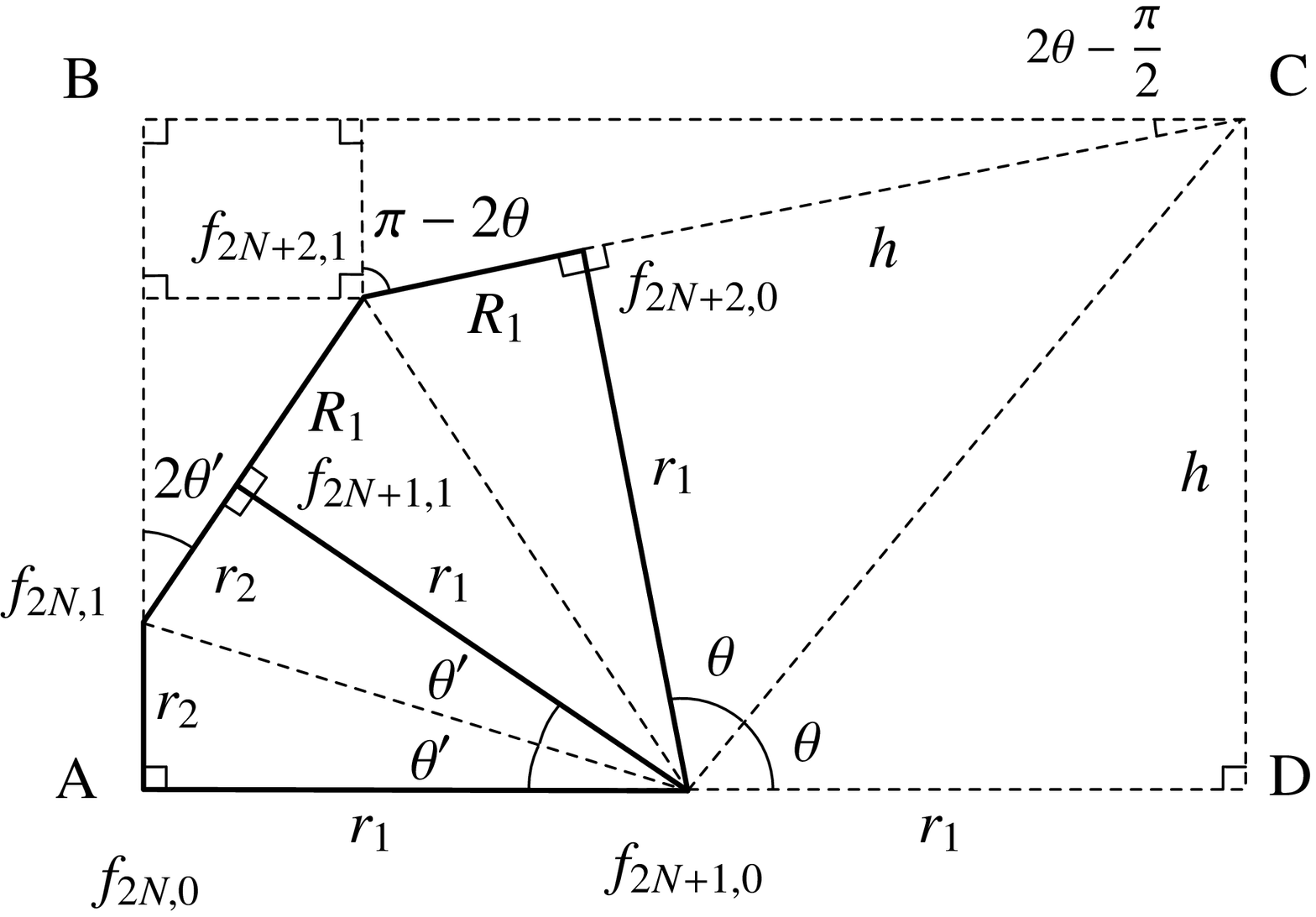}\\
(ii) $\frac{\pi}{2}\le2\theta<\pi$
\end{minipage}
\caption{{\small Configuration of points in Lemma \ref{angles}.}}\label{fig:delta_theta1}
\end{figure}
First, we consider the case of $0\le2\theta\le\pi/2$, see Figure \ref{fig:delta_theta1} (i). From ${\rm BC}={\rm AD}$ we obtain
\begin{equation}
(r_2+R_1)\sin2\theta'+(h+R_1)\sin2\theta=2r_1,
\end{equation} 
which yields
\begin{equation}
(r_2h+r_1^2)\left[(r_2+R_1)h+(r_2R_1-r_1^2)\right]=0.
\end{equation} 
Since $r_2h+r_1^2>0$, we have $h=\dfrac{r_1^2-r_2R_1}{r_2+R_1}$ and
\begin{equation}\label{tan_theta}
\tan\theta=\dfrac{r_1^2-r_2R_1}{r_1(r_2+R_1)}.
\end{equation}
Thus we get (\ref{theta}) from (\ref{delta_R}). When $\pi/2\le2\theta<\pi$, the configuration of points is shown in Figure \ref{fig:delta_theta1} (ii). The equality ${\rm BC}={\rm AD}$ implies
\begin{equation}
(r_2+R_1)\sin2\theta'+(h+R_1)\sin(\pi-2\theta)=2r_1,
\end{equation} 
which gives the same result as the case of $0\le2\theta\le\pi/2$. Let us investigate the case of $-\pi/2\le2\theta\le0$, see Figure \ref{fig:delta_theta2} (i). Since ${\rm DC}={\rm AB}$ we have
\begin{equation}
(h+R_1)\cos2\theta+(r_2+R_1)\cos(\pi-2\theta')=h+r_2, 
\end{equation}
which leads us to 
\begin{equation}\label{eq_h}
(r_2h+r_1^2)\left[(r_2R_1-r_1^2)h-r_1^2(r_2+R_1)\right]=0. 
\end{equation}
Then we have $h=\dfrac{r_1^2(r_2+R_1)}{r_2R_1-r_1^2}$, and thus (\ref{tan_theta}). Figure \ref{fig:delta_theta2} (ii) illustrates the case of $-\pi<2\theta\le-\pi/2$. We see from ${\rm AB}={\rm DC}$ that
\begin{equation}
(h+R_1)\cos(\pi-|2\theta|)+h+r_2=(r_2+R_1)\cos(\pi-2\theta'),
\end{equation}
which also leads us to (\ref{eq_h}), and thus (\ref{tan_theta}). Therefore we have proved Lemma \ref{angles}.\qed

\medskip

\noindent
From (\ref{f_and_theta}), (\ref{theta}) and the initial condition $f_{0,0}=0,\,f_{1,0}=1$, 
we see by induction that the points $f_{n,0}$ satisfy 
\begin{equation}
\gamma f_{n,0}=2n
\dfrac{(f_{n+1,0}-f_{n,0})(f_{n,0}-f_{n-1,0})}{f_{n+1,0}-f_{n-1,0}}. 
\end{equation}
Similarly, we see that $f_{0,m}$ satisfy 
\begin{equation}
\gamma f_{0,m}=2m
\dfrac{(f_{0,m+1}-f_{0,m})(f_{0,m}-f_{0,m-1})}{f_{0,m+1}-f_{0,m-1}}. 
\end{equation}
Thus it is possible to determine $f_{n,m}$ in $\BZ_+^2$ by using
(\ref{cr}).  Since (\ref{cr}) is compatible with (\ref{eq-f}), $f_{n,m}$
satisfies (\ref{eq-f}) simultaneously. This proves the second part of
Theorem \ref{thm:immersion_R}.\qed

\subsection{Positivity of radii of circles}
Theorem \ref{thm:immersion_R} claims that if $R_{n,m}$ satisfying (\ref{eq_1}) and (\ref{eq_2}) is
positive, then the corresponding $f_{n,m}$ is an immersion. In order to establish Theorem
\ref{thm:immersion}, we have to prove the positivity of $R_{n,m}$ determined by (\ref{eq_1}) and
(\ref{eq_2}) with the initial condition $R_{1,0}=1$, $R_{0,1}=\zeta\,(>0)$. First, we show that
positivity of $R_{m,n}$ for $(n,m)\in{\bf V}$ is reduced to that of $R_{n,n+1}$ for
$n\in\mathbb{Z}_+$.

\begin{prop}\label{positivity}
Let the solution $R_{n,m}$ of (\ref{eq_1}) for $(n,m)\in{\bf V}$ and (\ref{eq_2}) for $n=m\in\BZ_+$
with initial data
\begin{equation}
R_{1,0}=1,\quad R_{0,1}=\zeta\,(>0)\label{init_R}
\end{equation}
be positive for $m=n+1$, namely, $R_{n,n+1}>0$ for any $n\in\BZ_+$. Then
$R_{n,m}$ is positive everywhere in ${\bf V}$ and satisfies (\ref{eq_2})
for $(n,m)\in{\bf V}$.
\end{prop}

\noindent \textbf{Proof.}\quad Equation (\ref{eq_2}) for $(n,m)=(0,0)$ with initial data
(\ref{init_R}) determines $R_{1,2}$. We use (\ref{eq_1}) and (\ref{eq_2}) for $n=m$ inductively to
get $R_{n,n+1}$ and $R_{n+1,n}$. As was mentioned before, we see that $R_{2n+1,0}=1$ and
$R_{0,2m+1}=\zeta$ for all $n,m\in\BZ_+$ by putting $n=0$ and $m=0$, respectively, in
(\ref{eq_1}). With these data one can determine $R_{n,m}$ in ${\bf V}$ by using (\ref{eq_1}). When $n\ge m$, we
use (\ref{eq_1}) in the form of
\begin{equation}
R_{n+1,m}=R_{n-1,m}
\dfrac{(n-m)R_{n,m+1}+(n+m)R_{n,m-1}}{(n+m)R_{n,m+1}+(n-m)R_{n,m-1}}. 
\end{equation}
For positive $R_{n-1,m},R_{n,m+1}$ and $R_{n,m-1}$, we get $R_{n+1,m}>0$. When $m\ge n$, one can
show in a similar way that $R_{n,m+1}>0$ for given positive $R_{n,m-1},R_{n+1,m}$ and $R_{n-1,m}$ by
using (\ref{eq_1}). One can show by induction that we have (\ref{(2,-1)}) for $m=n\ge1$, and we
get (\ref{eq_2}) for $n=m+2$. Similarly, one can show by induction that we have (\ref{(2,-1)}) at
$(n+2k,n)$ for $n,k\ge1$, and (\ref{eq_2}) for $(n+2k,n)$. Thus we obtain (\ref{eq_2}) for $n\ge m$. One can show in a similar way that we have (\ref{eq_2}) for $n\le m$ by using (\ref{(2,-1)}) at $(n,n+2k)$ as an auxiliary relation. \qed
\par\bigskip

Due to Proposition \ref{positivity}, the discrete function $Z^{\gamma}$ with ${\rm Re}~\gamma=1$ is an immersion if and only if $R_{n,n+1}>0$ for all $n\in\BZ_+$. We next reduce the positivity to the existence of unitary solution to a certain system of difference equations.

\begin{prop}\label{prop:xy}
The map $f\,:\,\BZ_+^2\to\BC$ satisfying (\ref{cr}) and (\ref{eq-f}) with the initial condition
$f_{0,0}=0,f_{1,0}=1,f_{0,1}=i\zeta\,(\zeta>0)$ is an immersion if and only if the solution
$(x_n,y_n)$ to the system of equations
\begin{equation}
\begin{array}{l}
(x_n-1)\dfrac{y_{n+1}-1}{x_n+y_{n+1}}
+(x_n+1)\dfrac{y_n-1}{x_ny_n-1}=0,\\[4mm]
\dfrac{\gamma}{2}
=\dfrac{n+2}{1-x_{n+1}^{-1}y_{n+1}^{-1}}
-\dfrac{n+1}{1+x_n^{-1}y_{n+1}},\quad x_0y_0=\dfrac{\gamma}{\gamma-2},
\end{array}   \label{alpha&varphi}
\end{equation}
with 
\begin{equation}
y_0=\dfrac{\zeta+i}{\zeta-i}, 
\end{equation}
is of the form $x_n=e^{2i\aa_n},y_n=e^{2i\varphi_n}$, where $\aa_n,\varphi_n\in(0,\pi/2)$. 
\end{prop}

\begin{figure}
\centering
\includegraphics[scale=0.4]{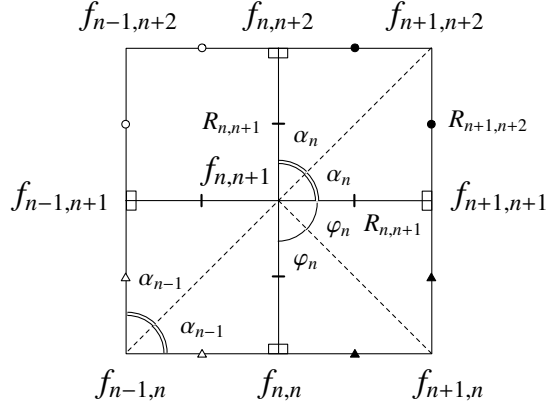}
\caption{{\small A schematic diagram of quadrilaterals for Proposition \ref{prop:xy}.}}
\label{fig:f_configuration}
\end{figure}

\noindent
\textbf{Proof.}\quad Let $f_{n,m}$ be an immersion. Define $\aa_n,\varphi_n\in(0,\pi/2)$ through 
\begin{equation}\label{alpha_varphi}
f_{n,n+2}-f_{n,n+1}=e^{2i\aa_n}(f_{n+1,n+1}-f_{n,n+1}),\quad
f_{n+1,n+1}-f_{n,n+1}=e^{2i\varphi_n}(f_{n,n}-f_{n,n+1}).  
\end{equation}
Using Proposition \ref{prop:kite}, one obtains
\begin{equation}
\begin{array}{ll}
f_{n+1,n+1}-f_{n,n+1}=R_{n,n+1}e^{i(2\varphi_n-\pi/2+\beta_n)},&
f_{n,n+1}-f_{n-1,n+1}=R_{n,n+1}e^{i(\beta_n+2\aa_{n-1}-\pi/2)},\\[2mm]
f_{n,n+2}-f_{n,n+1}=R_{n,n+1}e^{i(2\varphi_n-\pi/2+2\aa_n+\beta_n)},&
f_{n,n+1}-f_{n,n}=R_{n,n+1}e^{i(\beta_n+\pi/2)},
\end{array}
\end{equation}
where $\beta_n=\arg\,(f_{n+1,n}-f_{n,n})$. Figure \ref{fig:f_configuration} is a schematic diagram illustrating the configuration of the relevant quadrilaterals.
Now the constraint (\ref{eq-f}) for $(n,n+1)$ is equivalent to
\begin{equation}
\gamma f_{n,n+1}=-2iR_{n,n+1}e^{i\beta_n}
\left(\dfrac{n}{e^{-2i\varphi_n}+e^{-2i\aa_{n-1}}}
+\dfrac{n+1}{e^{-2i(\varphi_n+\aa_n)}-1}\right). 
\end{equation}
Putting these expressions into the equality
\begin{equation}
f_{n+1,n+2}-f_{n,n+1}
=e^{i(2\varphi_n-\pi/2+\beta_n)}(R_{n,n+1}+iR_{n+1,n+2}),
\end{equation}
together with $R_{n+1,n+2}=R_{n,n+1}\tan\aa_n$ and $e^{i\beta_{n+1}}=e^{i(\beta_n-\pi/2+2\varphi_n)}$, one obtains 
\begin{equation}
\begin{array}{l}
2i\sin\aa_n\left(\dfrac{n+1}{e^{-2i\varphi_{n+1}}+e^{-2i\aa_n}}
+\dfrac{n+2}{e^{-2i(\varphi_{n+1}+\aa_{n+1})}-1}\right)\\[4mm]
+2\cos\aa_n\left(\dfrac{n}{1+e^{2i(\varphi_n-\aa_{n-1})}}
+\dfrac{n+1}{e^{-2i\aa_n}-e^{2i\varphi_n}}\right)=-\gamma e^{i\aa_n}.
\end{array}   \label{alpha&varphi'}
\end{equation}
On the other hand, equation (\ref{eq_2}) for $m=n$ is reduced to 
\begin{equation}
\delta R_{n,n+1}
=(n+1)\dfrac{R_{n+1,n}R_{n+1,n+2}-R_{n,n+1}^2}{R_{n+1,n}+R_{n+1,n+2}}
-n\dfrac{R_{n,n+1}^2-R_{n+1,n}R_{n-1,n}}{R_{n+1,n}+R_{n-1,n}}.
\end{equation}
By a similar geometric consideration to the proof of Lemma \ref{angles}, this implies
\begin{equation}
\delta=-(n+1)\cot(\aa_n+\varphi_n)-n\tan(\aa_{n-1}-\varphi_n),
\end{equation}
which can be transformed to
\begin{equation}
\dfrac{\gamma}{2}
=\dfrac{n+1}{1-e^{-2i(\aa_n+\varphi_n)}}
-\dfrac{n}{1+e^{2i(\varphi_n-\aa_{n-1})}}.
\label{alpha&varphi''}
\end{equation}
By using (\ref{alpha&varphi''}), we see that (\ref{alpha&varphi'}) yields the first equation of
(\ref{alpha&varphi}) with $x_n=e^{2i\aa_n}$ and $y_n=e^{2i\varphi_n}$. The second equations of
(\ref{alpha&varphi}) come from (\ref{alpha&varphi''}). This proves the necessity part.

\medskip

\noindent Now let us suppose that there is a solution $(x_n,y_n)=(e^{2i\aa_n},e^{2i\varphi_n})$ of (\ref{alpha&varphi}) with $\aa_n,\varphi_n\in(0,\pi/2)$. This solution together with (\ref{alpha_varphi}) and (\ref{cr}) determines a sequence of orthogonal circles with their centers on $f_{n,n+1}$, and thus the points $(f_{n,n+1},f_{n\pm1,n+1},f_{n,n},f_{n,n+2})$. Now (\ref{cr}) determines $f_{n,m}$ on $\BZ_+^2$. Since $\aa_n,\varphi_n\in(0,\pi/2)$, the inner parts of the quadrilaterals $(f_{n,n+1},f_{n+1,n+1},f_{n+1,n+2},f_{n,n+2})$ and of the quadrilaterals $(f_{n,n},f_{n+1,n},f_{n+1,n+1},f_{n,n+1})$ are disjoint, which means that we have positive solution $R_{n,n+1}$ and $R_{n+1,n}$ of (\ref{eq_1}) and (\ref{eq_2}). Given $R_{n+1,n}$ and $R_{n,n+1}$, (\ref{eq_1}) determines $R_{n,m}$ for all $(n,m)\in{\bf V}$. Due to Proposition \ref{positivity}, $R_{n,m}$ is positive, and satisfies (\ref{eq_1}) and (\ref{eq_2}). Theorem \ref{thm:immersion_R}
implies that the discrete conformal map $g_{n,m}$ corresponding to the circle pattern $\{C_{n,m}\}$ determined by $R_{n,m}$ is an immersion and satisfies (\ref{eq-f}). Since $g_{n,n}=f_{n,n}$ and $g_{n,n\pm1}=f_{n,n\pm1}$, equation (\ref{cr}) implies $g_{n,m}=f_{n,m}$. This proves Proposition \ref{prop:xy}. \qed

\medskip

\noindent Note that although (\ref{alpha&varphi}) is a system of equations, a solution $(x_n,y_n)$
of (\ref{alpha&varphi}) is determined by its initial value $y_0$.

\medskip

\noindent
The system of equation (\ref{alpha&varphi}) can be written in the following recurrent form: 
\begin{equation}\label{eqn:xy}
y_{n+1}=\dfrac{1+x_n^2-2x_ny_n}{x_n^2y_n+y_n-2x_n},\quad
x_{n+1}=\dfrac{(2n+2+\gamma)x_n+\gamma y_{n+1}}
              {y_{n+1}\big[(\gamma-2)x_n+(\gamma-2n-4)y_{n+1}\big]},
\end{equation}
or
\begin{equation}
y_{n+1}=\Phi(x_n,y_n),\quad x_{n+1}=\Psi(n,x_n,y_{n+1}), 
\end{equation}
\begin{equation}
\Phi(x,y)=y\dfrac{x^{-1}y^{-1}+xy^{-1}-2}{xy+x^{-1}y-2},\quad
\Psi(n,x,y)=-\dfrac{1}{x}
\dfrac{(2n+3+i\delta)xy^{-1}+(1+i\delta)}
      {(2n+3-i\delta)x^{-1}y+(1-i\delta)},
\end{equation}
and $x_0=-\dfrac{1+i\delta}{1-i\delta}\dfrac{1}{y_0}$. It is easy to see that
$|\Phi(x,y)|=|\Psi(n,x,y)|=1$ when $|x|=|y|=1$ and that $|x_0|=1$ when $|y_0|=1$, which implies that
this system possesses unitary solutions. Moreover, we have the following theorem as for the arguments
of the unitary solutions:
\begin{thm}
There exists a unitary solution $(x_n,y_n)$ to the system of equation (\ref{alpha&varphi}) with
$x_n,y_n\in A_{\rm I}\bs\{\pm1\}$, where
\begin{equation}
A_{\rm I}=\{e^{2i\beta}\,|\,\beta\in[0,\pi/2]\}. 
\end{equation}
\end{thm}

\medskip

\noindent \textbf{Proof}\quad We first investigate the properties of the function $\Phi(x,y)$ and
$\Psi(n,x,y)$ restricted to the torus $T^2=S^1\times S^1=\{(x,y)\,|\,x,y\in\BC,\,|x|=|y|=1\}$.

\medskip

\noindent \textbf{Property 1.} The function $\Phi(x,y)$ is continuous on $A_{\rm I}\times A_{\rm
I}\bs\{(\pm1,\pm1)\}$. The function $\Psi(n,x,y)$ is continuous on $A_{\rm I}\times A_{\rm I}$ for
any $n\in\BZ_+$. (Continuity on the boundary of $A_{\rm I}\times A_{\rm I}$ is understood to be
one-sided.)

\medskip

\noindent The points of discontinuity must satisfy
\begin{equation}
x^2y+y-2x=0,\quad (2n+3-i\delta)y+(1-i\delta)x=0. 
\end{equation}
The first identity holds only for $(x,y)=(\pm1,\pm1)$. The second never holds for unitary $x,y$.

\medskip

\noindent \textbf{Property 2.} For $(x,y)\in A_{\rm I}\times A_{\rm I}\bs\{(\pm1,\pm1)\}$, we have
$\Phi(x,y)\in A_{\rm I}$. For $(x,y)\in A_{\rm I}\times A_{\rm I}$, we have $\Psi(n,x,y)\in A_{\rm
I}\cup A_{\rm II}\cup A_{\rm IV}$, where $A_{\rm II}:=\{e^{2i\beta}\,|\,\beta\in(\pi/2,\pi]\}$ and
$A_{\rm IV}:=\{e^{2i\beta}\,|\,\beta\in[-\pi/2,0)\}$.

\medskip

\noindent Property 2 is verified as follows: using the transformation 
\begin{equation}
u_n=\tan\aa_n,\quad v_n=\tan\varphi_n, 
\end{equation}
where $x_n=e^{2i\alpha_n} $and $y_n=e^{2i\varphi_n}$, we see that the first equation of
(\ref{alpha&varphi}) takes the form
\begin{equation}
v_{n+1}=u_n^{-2}v_n.\label{ev:v}
\end{equation}
It is obvious that $u^{-2}v\in[0,+\infty]$ when
$(u,v)\in[0,+\infty]\times[0,+\infty]\bs\{(0,0),(+\infty,+\infty)\}$. The second equation of (\ref{alpha&varphi}) can be expressed as
\begin{equation}
\aa_{n+1}=\omega_n-\aa_n+\dfrac{\pi}{2},\quad
e^{2i\omega_n}=
\dfrac{(2n+3+i\delta)x_ny_{n+1}^{-1}+(1+i\delta)}
      {(2n+3-i\delta)x_n^{-1}y_{n+1}+(1-i\delta)}.\label{ev:alpha}
\end{equation}
By using the variables $u_n,v_{n+1}$, we have 
\begin{equation}
\tan(\omega_n-\aa_n)=F(n,u_n,v_{n+1}), 
\end{equation}
where 
\begin{equation}
F(n,u,v)=\dfrac{[(n+1)-(n+2)v^2]u-(2n+3)v+\delta(1+uv)}
               {(n+2)-(n+1)v^2+(2n+3)uv+\delta v(1+uv)}. 
\end{equation}

\begin{lem}\label{lem:alpha_n+1}
It holds that $\omega_n-\aa_n+\frac{\pi}{2}\in[-\pi/2,\pi]$ for $\aa_n,\varphi_{n+1}\in[0,\pi/2]$. 
\end{lem}

\begin{figure}
 \centering
\includegraphics[scale=0.4]{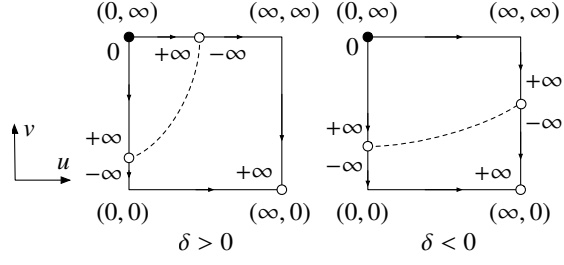}
\caption{{\small Behaviour of $F(n,u,v)$.}}
\label{fig:Fuv}
\end{figure}

\noindent
\textbf{Proof.}\quad Let us investigate the function $F(n,u,v)$ for $u,v\in[0,+\infty]$. It is easy to see that 
\begin{equation}
\dfrac{\partial F(n,u,v)}{\partial u}>0,\quad
\dfrac{\partial F(n,u,v)}{\partial v}<0
\end{equation}
on $[0,+\infty]^2$ except for the points satisfying $(n+2)-(n+1)v^2+(2n+3)uv+\delta v(1+uv)=0$. Consider the values of $F(n,u,v)$ on the boundary of $[0,+\infty]^2$. It is easy to see that
\begin{equation}
F(n,0,+\infty)=0,\quad F(n,0,0)=\dfrac{\delta}{n+2},\quad
F(n,+\infty,+\infty)=-\dfrac{n+2}{\dl},\quad F(n,+\infty,0)=+\infty.
\end{equation}
We find that $\dfrac{dF(n,0,v)}{dv}<0$ except for the point
$v=v_+:=\dfrac{\delta+\sqrt{\delta^2+4(n+1)(n+2)}}{2(n+1)}$, and that
$F(n,u,0)=\dfrac{(n+1)u+\delta}{n+2}$ is monotone increasing. Note that
\begin{equation}
F(n,u,+\infty)=\dfrac{(n+2)u}{n+1-\delta u},\quad
F(n,+\infty,v)=\dfrac{-(n+2)v^2+\delta v+n+1}{v(\delta v+2n+3)}.
\end{equation}
When $\delta>0$, we see that $\dfrac{dF(n,u,+\infty)}{du}>0$ except for the point
$u=\dfrac{n+1}{\delta}$ and that $\dfrac{dF(n,+\infty,v)}{dv}<0$ except for the point $v=0$. When
$\delta<0$, we see that $\dfrac{dF(n,u,+\infty)}{du}>0$ and that $\dfrac{dF(n,+\infty,v)}{dv}<0$
except for the points $v=0,v_*:=-\dfrac{2n+3}{\delta}$. The singular points of $F(n,u,v)$ can be
expressed by $(u,v)=(G(v),v)$, where $G(v)=\dfrac{(n+1)v^2-\delta v-(n+2)}{v(\delta v+2n+3)}$, if
$v\ne0,v_*$. Note that $(u,v)\in[0,+\infty]^2$. Then we find that the singular points lie in
$v\in[v_+,+\infty],u\in[0,(n+1)/\dl]$ (when $\delta>0$) or in $v\in[v_+,v_*],u\in[0,+\infty]$ (when
$\delta<0$), and that $G(v)$ is monotone increasing (See Figure \ref{fig:Fuv}). Therefore we see that
$\omega_n-\aa_n\in[-\pi,\pi/2]$ when $\aa_n,\varphi_{n+1}\in[0,\pi/2]$. \qed

\medskip

\noindent
The final equation in (\ref{alpha&varphi}) leads us to 
\begin{equation}
\aa_0=-\varphi_0+\theta+\dfrac{\pi}{2},\quad\tan\theta=\delta. 
\end{equation}
It is easy to see that $-\varphi_0+\theta+\pi/2\in[-\pi/2,\pi]$ when
$\varphi_0\in(0,\pi/2)$. Therefore property 2 is established.

\medskip

\noindent
Now let us introduce
\begin{equation}
\begin{array}{l}
S_{\rm II}(k):=\{y_0\in A_{\rm I}\,|
\,x_k\in A_{\rm II},\,x_l\in A_{\rm I}\,(l<k)\},
\\[1mm]
S_{\rm IV}(k):=\{y_0\in A_{\rm I}\,|
\,x_k\in A_{\rm IV},\,x_l\in A_{\rm I}\,(l<k)\},
\end{array} 
\end{equation}
where $(x_n,y_n)$ is the solution of (\ref{alpha&varphi}). 
From property 1, it follows that $S_{\rm
II}(k)$ and $S_{\rm IV}(k)$ are open sets in the induced topology of $A_{\rm I}$. Denote
\begin{equation}
S_{\rm II}=\cup S_{\rm II}(k),\quad S_{\rm IV}=\cup S_{\rm IV}(k), 
\end{equation}
which are also open. These sets are nonempty since $S_{\rm II}(1)$ and $S_{\rm IV}(1)$ are nonempty
(See the proof of Lemma \ref{lem:alpha_n+1}). Note that $S_{\rm II}(0)$ or $S_{\rm IV}(0)$ can be
empty.
%
Finally, introduce
\begin{equation}
S_{\rm I}:=\{y_0\in A_{\rm I}\,|\,x_n\in A_{\rm I}\,\,
\mbox{for all}\,\,n\in\BZ_+\}. 
\end{equation}
It is obvious that $S_{\rm I},S_{\rm II}$ and $S_{\rm IV}$ are mutually disjoint. Property 2 implies 
\begin{equation}
S_{\rm I}\cup S_{\rm II}\cup S_{\rm IV}=A_{\rm I}. 
\end{equation}
Since the connected set $A_{\rm I}$ cannot be covered by two open disjoint subsets $S_{\rm II}$ and
$S_{\rm IV}$, we see that $S_{\rm I}\ne\emptyset$.  So there exists $y_0$ such that the solution
$(x_n,y_n)\in A_{\rm I}\times A_{\rm I}$ for any $n\in\BZ_+$. Suppose that $\aa_n=0,\varphi_n\ne0$
hold at a certain $n$. Then we get $\varphi_{n+1}=\pi/2$ and $\aa_{n+1}=-\pi/2$ from
(\ref{eqn:xy}), or (\ref{ev:v}) and (\ref{ev:alpha}). Similarly, if $\aa_n=\pi/2,\varphi_n\ne\pi/2$
hold at a certain $n$, we get $\varphi_{n+1}=0$ and then $\aa_{n+1}=\pi$. It means that in both
cases $x_{n+1}\not\in A_{\rm I}$. Suppose that $\varphi_n\ne0,\varphi_{n+1}=0$ hold at a certain
$n$. Then we get $\aa_n=\pi/2$ and then $\aa_{n+1}=\pi$ from (\ref{eqn:xy}), or (\ref{ev:v}) and
(\ref{ev:alpha}). Similarly, if $\varphi_n\ne\pi/2,\varphi_{n+1}=\pi/2$ hold at a certain $n$, we
get $\aa_n=0$ and then $\aa_{n+1}=-\pi/2$. It also means that in both cases $x_{n+1}\not\in A_{\rm
I}$. Thus, it follows that $\aa_n\ne0,\pi/2$ and $\varphi_n\ne0,\pi/2$ for the solution
$(x_n,y_n)\in A_{\rm I}\times A_{\rm I}$ for any $n\in\BZ_+$. \qed

\medskip

We have shown that $S_{\rm I}$ is not empty. In order to establish Theorem \ref{thm:immersion}, let
us finally show the uniqueness of the initial condition which gives rise to the solution
$(x_n,y_n)\in A_{\rm I}\times A_{\rm I}\bs \{(\pm1,\pm1)\}$, namely, the circle pattern. Indeed, the
initial condition is nothing but that for the discrete power function.  Take a solution $(x_n,y_n)$
such that $y_0\in S_{\rm I}$ and consider the corresponding circle pattern. Let $R_k$ be radii of
circles with centers at $f_{2k,1}$, i.e., $R_k:=R_{2k,1}$. We have the following lemma.

\begin{lem}
An explicit formula for $R_k$ is given by 
\begin{equation}
R_k=i\,
\dfrac{(-1)^k
\left[\left(\frac{r-k+1}{2}\right)_k-\left(\frac{r-k}{2}\right)_k\right]-i\zeta
\left[\left(\frac{r-k+1}{2}\right)_k+\left(\frac{r-k}{2}\right)_k\right]}
{(-1)^k
\left[\left(\frac{r-k+1}{2}\right)_k+\left(\frac{r-k}{2}\right)_k\right]-i\zeta
\left[\left(\frac{r-k+1}{2}\right)_k-\left(\frac{r-k}{2}\right)_k\right]}, 
\end{equation}
where $\zeta=R_0$. 
\end{lem}

\medskip

\noindent \textbf{Proof}\quad The radii $R_k$ are defined by
$R_k=|f_{2k,0}-f_{2k,1}|=|v_{2k,0}||v_{2k,1}|$ (see (\ref{f-v})). From
Proposition \ref{formula_v}, we have
\begin{equation}
v_{2k,0}=\dfrac{(r)_k}{(-r+1)_{k}},\quad
v_{2k,1}=\dfrac{\varphi(-k+1,-r-k+1,-r+2;-1)}{\varphi(-k,-r-k+1,-r+1;-1)}, 
\end{equation}
where $\varphi(a,b,c;-1)$ is given by (see (\ref{Gauss}))
\begin{equation}
\begin{split}
\varphi(a,b,c;-1)
&=\dfrac{\Gamma(a)\Gamma(b)}{\Gamma(c)}F(a,b,c;-1)\\
&+c_1e^{\pi i(1-c)}\dfrac{\Gamma(a-c+1)\Gamma(b-c+1)}{\Gamma(2-c)}
F(a-c+1,b-c+1,2-c;-1),   
\end{split}
\end{equation}
and 
\begin{equation}
\zeta=c_1e^{-\pi\delta/2}\,(>0) .
\end{equation}
It is easy to see that $|v_{2k,0}|=1$ under
${\rm Re}\,\gamma=1$. Due to the formulae\,\cite{AS}
\begin{align}
&F(a,b,a-b+1;-1)=2^{-a}\pi^{1/2}\dfrac{\Gamma(a-b+1)}{\Gamma(\hf a-b+1)\Gamma(\hf a+\hf)},\\[4mm]
\begin{split}
&F(a,b,a-b+2;-1)=2^{-a}\pi^{1/2}(b-1)^{-1}\Gamma(a-b+2)\\
&\hskip40mm\times\left[\dfrac{1}{\Gamma(\hf a)\Gamma(\hf a-b+\frac{3}{2})}
-\dfrac{1}{\Gamma(\hf a+\hf)\Gamma(\hf a-b+1)}\right],
\end{split}
\end{align}
and the contiguity relation
\begin{equation}
 \begin{split}
&(c-a)(c-b)tF(a,b,c+1;t)+c\left[(a+b-2c+1)t+c-1\right]F(a,b,c;t)\\
&\hskip40mm +c(c-1)(t-1)F(a,b,c-1;t)=0,  
 \end{split}
\end{equation}
we get 
\begin{equation}
v_{2k,1}=
\dfrac{(-1)^k
\left[\left(\frac{r-k+1}{2}\right)_k-\left(\frac{r-k}{2}\right)_k\right]-i\zeta
\left[\left(\frac{r-k+1}{2}\right)_k+\left(\frac{r-k}{2}\right)_k\right]}
{(-1)^k
\left[\left(\frac{r-k+1}{2}\right)_k+\left(\frac{r-k}{2}\right)_k\right]-i\zeta
\left[\left(\frac{r-k+1}{2}\right)_k-\left(\frac{r-k}{2}\right)_k\right]}. 
\end{equation}
Then we can easily verify that $v_{2k,1}$ satisfy the recurrence relation 
\begin{equation}
v_{2(k+1),1}=-\dfrac{1-i\ep_kv_{2k,1}}{v_{2k,1}-i\ep_k},\quad v_{0,1}=-i\zeta,
\end{equation}
where $\ep_k=\dfrac{\delta}{2k+1}$. On the other hand, (\ref{(2,1)}) for $m=0$ is reduced to 
\begin{equation}
R_{k+1}=\dfrac{1-\ep_kR_k}{R_k+\ep_k}.
\end{equation}
Then noticing $R_0=\zeta$, we see that $R_k=|v_{2k,1}|=iv_{2k,1}$. \qed

\begin{prop}\label{unique}
The set $S_{\rm I}$ consists of only one element $\dfrac{e^{-\pi\delta/2}+i}{e^{-\pi\delta/2}-i}$. 
\end{prop}

\medskip

\noindent
\textbf{Proof}\quad By using the formulae
\begin{equation}
\Gamma(z)\Gamma(1-z)=\dfrac{\pi}{\sin\pi z},
\end{equation}
and the asymptotic formula as $z\to\infty$
\begin{equation}
\Gamma(az+b)\sim(2\pi)^{1/2}e^{-az}(az)^{az+b-\hf}\quad
(|\arg z|<\pi\,,\ a>0), 
\end{equation}
we get 
\begin{equation}\label{eqn:lim_R}
\lim_{l\to\infty}R_{2l}
=\dfrac{(1-e^{-\pi\delta/2})+\zeta(1+e^{-\pi\delta/2})}
       {(1+e^{-\pi\delta/2})-\zeta(1-e^{-\pi\delta/2})},\quad
\lim_{l\to\infty}R_{2l-1}
=\dfrac{(1+e^{-\pi\delta/2})-\zeta(1-e^{-\pi\delta/2})}
       {(1-e^{-\pi\delta/2})+\zeta(1+e^{-\pi\delta/2})}. 
\end{equation}
\begin{figure}
 \centering
\includegraphics[scale=0.4]{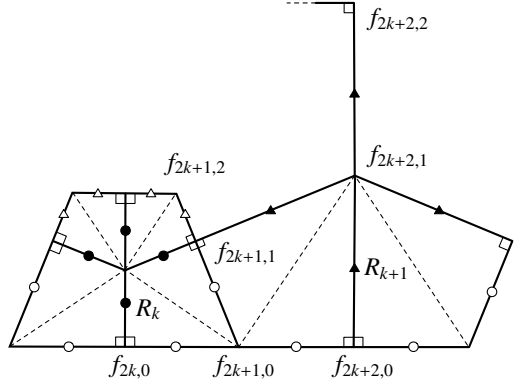}
\caption{{\small Configuration of points around $f_{2k,0}$ for large $k$.}}
\label{fig:f_asymptotic}
\end{figure}
The relation (\ref{theta}) implies $\disp\lim_{N\to\infty}\theta_N=0$. Figure \ref{fig:f_asymptotic} illustrates the
configuration around $f_{2k,0}$ for sufficiently large $k$. Note that the three points $f_{2k,0}$,
$f_{2k+1,0}$, $f_{2k+2,0}$ are asymptotically collinear. We see from (\ref{eqn:lim_R}) that
$R_k$ should behave as $\disp\lim_{k\to\infty}R_{k}=1$ irrespective of the parity of $k$, since the
four points $f_{2k+1,1},f_{2k+2,1},f_{2k+2,2},f_{2k+1,2}$ form a quadrilateral. Then we find that
$\zeta=e^{-\pi\delta/2}$. This completes the proof of Proposition \ref{unique}. \qed
\par\bigskip

Therefore $f_{n,m}$ satisfying (\ref{cr}) and (\ref{eq-f}) is an immersion 
if and only if the initial condition is given by (\ref{initial_f_immsesion}). This completes
the proof of Theorem \ref{thm:immersion}.

\section{Concluding remarks}
The discrete logarithmic function and cases where $\gamma\in 2\mathbb{Z}$ were excluded from the
considerations in the previous sections. From the viewpoint of the theory of hypergeometric
functions, these cases lead to integer differences in the characteristic exponents. Thus we need a
different treatment for the precise description of these cases. However, they may be obtained by some
limiting procedures in principle. In fact, Agafonov has examined the case where $\gamma=2$ and
$\gamma=0$ by using a limiting procedure\,\cite{Agafonov_1,Agafonov_2}, the former is the discrete
power function $Z^2$ and latter is the discrete logarithmic function. In general, one may obtain a
description of these cases by introducing the functions $\widetilde{f}_{n,m}$ and
$\widehat{f}_{n,m}$ as
\begin{equation}
\widetilde{f}_{n,m}:=\left\{
\begin{array}{ll}
\displaystyle\lim_{r\to j}
\dfrac{1}{j}\dfrac{(-r+1)_j}{(r+1)_{j-1}}f_{n,m},
&\mbox{for $\gamma=2j\in 2\mathbb{Z}_{>0}$}\\[4mm]
\displaystyle\lim_{r\to-j}
\dfrac{(-r+1)_j}{(r+1)_j}f_{n,m},
&\mbox{for $\gamma=-2j\in 2\mathbb{Z}_{<0}$}
\end{array}
\right.
\end{equation}
and 
\begin{equation}
\widehat{f}_{n,m}=\lim_{r\to0}\dfrac{f_{n,m}-1}{r}, 
\end{equation}
respectively. The function $\widetilde{f}_{n,m}$ might coincide with the counterpart defined in
section 6 of \cite{Agafonov-Bobenko:2000}.

Moreover, it has been shown that the discrete power function and logarithmic function associated
with hexagonal patterns are also described by some discrete Painlev\'e
equations\,\cite{Agafonov-Bobenko:2003}. It may be an interesting problem to construct the explicit
formula for them.

\section*{Acknowledgements} 
The authors would
like to express our sincere thanks to Professor Masaaki Yoshida for valuable suggestions and
discussions. This work was partially supported by JSPS Grant-in-Aid for Scientific Research
No. 21740126, 21656027 and 23340037, and by the Global COE Program Education and Research Hub for
Mathematics-for-Industry from the Ministry of Education, Culture, Sports, Science and Technology,
Japan.

\appendix
\makeatletter
\@addtoreset{equation}{section}
\renewcommand{\theequation}{\@Alph\c@section.\@arabic\c@equation}
\makeatother
\section{B\"acklund transformations of the sixth Painlev\'e equation\label{sym_P6}}
As a preparation, we give a brief review of the B\"acklund transformations and some of the bilinear
equations for the $\tau$ functions\,\cite{Masuda}. It is well-known that P$_{\rm VI}$ (\ref{P6}) is
equivalent to the Hamilton system
\begin{equation}
q'=\dfrac{\partial H}{\partial p},\quad p'=-\dfrac{\partial H}{\partial q},
\quad '=t(t-1)\dfrac{d}{dt},
\end{equation}
whose Hamiltonian is given by 
\begin{equation}
H=f_0f_3f_4f_2^2-[\alpha_4f_0f_3+\alpha_3f_0f_4+(\alpha_0-1)f_3f_4]f_2
+\alpha_2(\alpha_1+\alpha_2)f_0.
\end{equation}
Here $f_i$ and $\alpha_i$ are defined by 
\begin{equation}
f_0=q-t,\quad f_3=q-1,\quad f_4=q,\quad f_2=p,
\end{equation}
and
\begin{equation}
\alpha_0=\theta,\quad\alpha_1=\kappa_{\infty},\quad
\alpha_3=\kappa_1,\quad\alpha_4=\kappa_0
\end{equation}
with  $\alpha_0+\alpha_1+2\alpha_2+\alpha_3+\alpha_4=1$. The B\"acklund transformations of P$_{\rm VI}$ are described by 
\begin{equation}
s_i(\alpha_j)=\alpha_j-a_{ij}\alpha_i\quad (i,j=0,1,2,3,4),
\end{equation}
\begin{equation}
s_2(f_i)=f_i+\dfrac{\alpha_2}{f_2},\quad
s_i(f_2)=f_2-\dfrac{\alpha_i}{f_i}~~(i=0,3,4),
\end{equation}
\begin{equation}
\begin{array}{ll}
s_5:
&\alpha_0\leftrightarrow\alpha_1,\quad\alpha_3\leftrightarrow\alpha_4,\quad
f_2\mapsto -\dfrac{f_0(f_2f_0+\alpha_2)}{t(t-1)},\quad
f_4\mapsto t\dfrac{f_3}{f_0},\\[4mm]
s_6:
&\alpha_0\leftrightarrow\alpha_3,\quad\alpha_1\leftrightarrow\alpha_4,\quad
f_2\mapsto -\dfrac{f_4(f_4f_2+\alpha_2)}{t},\quad 
f_4\mapsto\dfrac{t}{f_4},\\[4mm]
s_7:
&\alpha_0\leftrightarrow\alpha_4,\quad\alpha_1\leftrightarrow\alpha_3,\quad
f_2\mapsto\dfrac{f_3(f_3f_2+\alpha_2)}{t-1},\quad 
f_4\mapsto\dfrac{f_0}{f_3},
\end{array}
\end{equation}
where $A=(a_{ij})_{i,j=0}^4$ is the Cartan matrix of type $D^{(1)}_4$. Then the group of birational
transformations $\langle s_0,\ldots,s_7\rangle$ generate the extended affine Weyl group
$\widetilde{W}(D^{(1)}_4)$. In fact, these generators satisfy the fundamental relations
\begin{equation}
s_i^2=1\quad(i=0,\ldots,7),\quad s_is_2s_i=s_2s_is_2\quad(i=0,1,3,4),
\end{equation}
and 
\begin{equation}
\begin{array}{c}
s_5s_{\{0,1,2,3,4\}}=s_{\{1,0,2,4,3\}}s_5,\quad
s_6s_{\{0,1,2,3,4\}}=s_{\{3,4,2,0,1\}}s_6,\quad
s_7s_{\{0,1,2,3,4\}}=s_{\{4,3,2,1,0\}}s_7,\\[1mm]
s_5s_6=s_6s_5,\quad s_5s_7=s_7s_5,\quad s_6s_7=s_7s_6.
\end{array}
\end{equation}

We add a correction term to the Hamiltonian $H$ as follows, 
\begin{equation}
H_0=H+\dfrac{t}{4}
\left[1+4\alpha_1\alpha_2+4\alpha_2^2-(\alpha_3+\alpha_4)^2\right]
+\dfrac{1}{4}
\left[(\alpha_1+\alpha_4)^2+(\alpha_3+\alpha_4)^2+4\alpha_2\alpha_4\right].
\end{equation}
This modification gives a simpler behavior of the Hamiltonian with respect to the B\"acklund
transformations. From the corrected Hamiltonian, we introduce a family of Hamiltonians
$h_i~(i=0,1,2,3,4)$ as
\begin{equation}
h_0=H_0+\frac{t}{4},\quad h_1=s_5(H_0)-\frac{t-1}{4},\quad 
h_3=s_6(H_0)+\frac{1}{4},\quad h_4=s_7(H_0), \quad h_2=h_1+s_1(h_1).
\label{def_h}
\end{equation}
Next, we also introduce $\tau$ functions $\tau_i~(i=0,1,2,3,4)$ by $h_i=(\log\tau_i)'$. Imposing the
condition that the action of the $s_i$'s on the $\tau$ functions also commute with the derivation
$'$, one can lift the B\"acklund transformations to the $\tau$ functions. The action of
$\widetilde{W}(D^{(1)}_4)$ is given by
\begin{equation}
s_0(\tau_0)=f_0\dfrac{\tau_2}{\tau_0},\quad
s_1(\tau_1)=\dfrac{\tau_2}{\tau_1},\quad
s_2(\tau_2)=\dfrac{f_2}{\sqrt{t}}
\dfrac{\tau_0\tau_1\tau_3\tau_4}{\tau_2},\quad
s_3(\tau_3)=f_3\dfrac{\tau_2}{\tau_3},\quad
s_4(\tau_4)=f_4\dfrac{\tau_2}{\tau_4},\label{act_on_tau}
\end{equation}
and 
\begin{equation}
\begin{array}{ll}
s_5:
&\tau_0\mapsto [t(t-1)]^{\frac{1}{4}}\tau_1,\quad
\tau_1\mapsto [t(t-1)]^{-\frac{1}{4}}\tau_0,\\[1mm]
&\tau_3\mapsto t^{-\frac{1}{4}}(t-1)^{\frac{1}{4}}\tau_4,\quad
\tau_4\mapsto t^{\frac{1}{4}}(t-1)^{-\frac{1}{4}}\tau_3,\quad
\tau_2\mapsto [t(t-1)]^{-\frac{1}{2}}f_0\tau_2,
\end{array}
\end{equation}
\begin{equation}
s_6:~
\tau_0\mapsto it^{\frac{1}{4}}\tau_3,\quad
\tau_3\mapsto -it^{-\frac{1}{4}}\tau_0,\quad
\tau_1\mapsto t^{-\frac{1}{4}}\tau_4,\quad
\tau_4\mapsto t^{\frac{1}{4}}\tau_1,\quad
\tau_2\mapsto t^{-\frac{1}{2}}f_4\tau_2,
\end{equation}
\begin{equation}
\begin{array}{ll}
s_7:
&\tau_0\mapsto (-1)^{-\frac{3}{4}}(t-1)^{\frac{1}{4}}\tau_4,\quad
\tau_4\mapsto(-1)^{\frac{3}{4}}(t-1)^{-\frac{1}{4}}\tau_0,\\[1mm]
&\tau_1\mapsto(-1)^{\frac{3}{4}}(t-1)^{-\frac{1}{4}}\tau_3,\quad
\tau_3\mapsto(-1)^{-\frac{3}{4}}(t-1)^{\frac{1}{4}}\tau_1,\\[1mm]
&\tau_2\mapsto -i(t-1)^{-\frac{1}{2}}f_3\tau_2. 
\end{array}
\end{equation}
We note that some of the fundamental relations are modified 
\begin{equation}
s_is_2(\tau_2)=-s_2s_i(\tau_2)\quad(i=5,6,7),
\end{equation}
and
\begin{equation}
\begin{array}{l}
s_5s_6\tau_{\{0,1,2,3,4\}}=\{i,-i,-1,-i,i\}s_6s_5\tau_{\{0,1,2,3,4\}},\\[1mm]
s_5s_7\tau_{\{0,1,2,3,4\}}=\{i,-i,-1,i,-i\}s_7s_5\tau_{\{0,1,2,3,4\}},\\[1mm]
s_6s_7\tau_{\{0,1,2,3,4\}}=\{-i,-i,-1,i,i\}s_7s_6\tau_{\{0,1,2,3,4\}}.
\end{array}
\end{equation}

Let us introduce the translation operators
\begin{equation}
\begin{array}{c}
\widehat{T}_{13}=s_1s_2s_0s_4s_2s_1s_7,\quad 
\widehat{T}_{40}=s_4s_2s_1s_3s_2s_4s_7,\\
\widehat{T}_{34}=s_3s_2s_0s_1s_2s_3s_5,\quad 
T_{14}=s_1s_4s_2s_0s_3s_2s_6,
\end{array}
\end{equation}
whose action on the parameters $\vec{\alpha}=(\alpha_0,\alpha_1,\alpha_2,\alpha_3,\alpha_4)$ is given by 
\begin{equation}
\begin{array}{c}
\widehat{T}_{13}(\vec{\alpha})=\vec{\alpha}+(0,1,0,-1,0),\\[1mm]
\widehat{T}_{40}(\vec{\alpha})=\vec{\alpha}+(-1,0,0,0,1),\\[1mm]
\widehat{T}_{34}(\vec{\alpha})=\vec{\alpha}+(0,0,0,1,-1),\\[1mm]
T_{14}(\vec{\alpha})=\vec{\alpha}+(0,1,-1,0,1).
\end{array}
\end{equation}
We denote
$\tau_{k,l,m,n'}=T_{14}^{n'}\widehat{T}_{34}^m\widehat{T}_{40}^l\widehat{T}_{13}^k(\tau_0)\,(k,l,m,n'\in\mathbb{Z})$. By
using this notation, we have
\begin{equation}
\begin{array}{ll}
\tau_{0,0,0,0}=\tau_0,&
\tau_{-1,-1,-1,0}=[t(t-1)]^{\frac{1}{4}}\tau_1,\\[1mm]
\tau_{0,-1,-1,0}=(-1)^{-\frac{3}{4}}t^{\frac{1}{4}}\tau_3,&
\tau_{0,-1,0,0}=(-1)^{-\frac{3}{4}}(t-1)^{\frac{1}{4}}\tau_4,\\[1mm]
\tau_{-1,-2,-1,1}=(-1)^{-\frac{1}{4}}s_0(\tau_0),&
\tau_{0,-1,0,1}=(-1)^{-\frac{3}{4}}[t(t-1)]^{\frac{1}{4}}s_1(\tau_1),\\[1mm]
\tau_{-1,-1,0,1}=-it^{\frac{1}{4}}s_3(\tau_3),& 
\tau_{-1,-1,-1,1}=(t-1)^{\frac{1}{4}}s_4(\tau_4), 
\end{array}   \label{lattice-tau}
\end{equation}
for instance. When the parameters $\vec{\alpha}$ take the values
\begin{equation}
(\alpha_0,\alpha_1,\alpha_2,\alpha_3,\alpha_4)=(-b,a+n',-n',c-a,b-c+1+n'), 
\end{equation}
the function $\tau_{k,l,m,n'}$ relates to the hypergeometric $\tau$ function $\tau_{n'}^{k,l,m}$ introduced in Proposition \ref{HG_P6} by\,\cite{Masuda}
\begin{equation}
\tau_{k,l,m,n'}=\omega_{k,l,m,n'}\tau_{n'}^{k,l,m}
t^{-(\hat{a}+\hat{b}-\hat{c}+2n')^2/4-(\hat{a}-\hat{b}-n')^2/4+n'(\hat{b}+n')-n'(n'-1)/2}
(t-1)^{(\hat{a}+\hat{b}-\hat{c}+2n')^2/4+1/2},
\end{equation}
where we denote $\hat{a}=a+k,\hat{b}=b+l+1$ and $\hat{c}=c+m$, and the constants
$\omega_{k,l,m,n'}=\omega_{k,l,m,n'}(a,b,c)$ are determined by the recurrence relations
\begin{equation}
\begin{array}{l}
\omega_{k+1,l,m,i}\omega_{k-1,l,m,i}
=i\hat{a}(\hat{c}-\hat{a})\omega_{k,l,m,i}^2, \\[1mm]
\omega_{k,l+1,m,i}\omega_{k,l-1,m,i}
=-i\hat{b}(\hat{c}-\hat{b})\omega_{k,l,m,i}^2,\\[1mm]
\omega_{k,l,m+1,i}\omega_{k,l,m-1,i}
=(\hat{c}-\hat{a})(\hat{c}-\hat{b})\omega_{k,l,m,i}^2
\end{array}   \quad (i=0,1)
\end{equation}
and 
\begin{equation}
\omega_{k,l,m,n'+1}\omega_{k,l,m,n'-1}=-\omega_{k,l,m,n'}^2
\end{equation}
with initial conditions 
\begin{equation}
\begin{array}{ll}
\omega_{-1,-2,-1,1}=(-1)^{-1/4}b,&\omega_{ 0,-2,-1,1}=b, \\
\omega_{-1,-1,-1,1}=1,&\omega_{ 0,-1,-1,1}=(-1)^{-1/4}, \\
\omega_{-1,0,0,1}=-(-1)^{-3/4}(c-a),&\omega_{ 0,0,0,1}=-i, \\
\omega_{-1,-1,0,1}=-i(c-a),&\omega_{ 0,-1,0,1}=(-1)^{-3/4}, 
\end{array}
\end{equation}
and 
\begin{equation}
\begin{array}{ll}
\omega_{-1,-2,-1,0}=(-1)^{-3/4}b,&\omega_{ 0,-2,-1,0}=-b, \\
\omega_{-1,-1,-1,0}=1,&\omega_{ 0,-1,-1,0}=(-1)^{-3/4}, \\
\omega_{-1,0,0,0}=(-1)^{-3/4}(c-a),&\omega_{ 0,0,0,0}=1, \\
\omega_{-1,-1,0,0}=c-a,&\omega_{ 0,-1,0,0}=(-1)^{-3/4}. 
\end{array}
\end{equation}

From the above formulation, one can obtain the bilinear equations for the $\tau$ functions. For
example, let us express the B\"acklund transformations
$s_2(f_i)=f_i+\dfrac{\alpha_2}{f_2}\,(i=0,3,4)$ in terms of the $\tau$ functions
$\tau_j\,(j=0,1,3,4)$. We have by using (\ref{act_on_tau})
\begin{equation}
\begin{array}{l}
\alpha_2t^{-\frac{1}{2}}\tau_3\tau_4
-s_1(\tau_1)s_2s_0(\tau_0)+s_0(\tau_0)s_2s_1(\tau_1)=0,\\[1mm]
\alpha_2t^{-\frac{1}{2}}\tau_0\tau_4
-s_1(\tau_1)s_2s_3(\tau_3)+s_3(\tau_3)s_2s_1(\tau_1)=0,\\[1mm]
\alpha_2t^{-\frac{1}{2}}\tau_0\tau_3
-s_1(\tau_1)s_2s_4(\tau_4)+s_4(\tau_4)s_2s_1(\tau_1)= 0.
\end{array}   \label{bi_BT}
\end{equation}
Applying the affine Weyl group $\widetilde{W}(D_4^{(1)})$ on these equations, we obtain 
\begin{equation}
\begin{array}{l}
(\alpha_0+\alpha_2+\alpha_4)\,t^{-\frac{1}{2}}\tau_3s_4(\tau_4)
-s_1(\tau_1)s_4s_2s_0(\tau_0)+\tau_0s_0s_4s_2s_1(\tau_1)=0,\\[1mm]
(\alpha_0+\alpha_2+\alpha_4)\,t^{\frac{1}{2}}\tau_1\tau_3
-\tau_4s_4s_2s_0(\tau_0)+\tau_0s_0s_2s_4(\tau_4)=0,
\end{array}   \label{bi_BT:1}
\end{equation}
\begin{equation}
\begin{array}{l}
(\alpha_0+\alpha_1+\alpha_2)\,t^{-\frac{1}{2}}\tau_3\tau_4
-\tau_1s_1s_2s_0(\tau_0)+\tau_0s_0s_2s_1(\tau_1)=0,\\[1mm]
(\alpha_0+\alpha_1+\alpha_2)\,t^{\frac{1}{2}}s_1(\tau_1)\tau_3
-s_4(\tau_4)s_1s_2s_0(\tau_0)+\tau_0s_0s_1s_2s_4(\tau_4)=0,
\end{array}   \label{bi_BT:2}
\end{equation}
\begin{equation}
\begin{array}{l}
(\alpha_2+\alpha_3+\alpha_4)\,t^{-\frac{1}{2}}\tau_0\tau_1
-\tau_4s_4s_2s_3(\tau_3)+\tau_3s_3s_2s_4(\tau_4)=0,\\[1mm]
(\alpha_2+\alpha_3+\alpha_4)\,t^{-\frac{1}{2}}s_4(\tau_4)\tau_0
-s_1(\tau_1)s_4s_2s_3(\tau_3)+\tau_3s_3s_4s_2s_1(\tau_1)=0,
\end{array}   \label{bi_BT:3}
\end{equation}
and
\begin{equation}
\begin{array}{l}
\alpha_2t^{-\frac{1}{2}}\tau_0\tau_3
-s_1(\tau_1)s_2s_4(\tau_4)+s_4(\tau_4)s_2s_1(\tau_1)=0,\\[1mm]
(\alpha_1+\alpha_4+\alpha_2)t^{-\frac{1}{2}}\tau_0\tau_3
-\tau_1s_1s_2s_4(\tau_4)+\tau_4s_4s_2s_1(\tau_1)=0.
\end{array}   \label{bi_BT:4}
\end{equation}
For instance, the first equation in (\ref{bi_BT:1}) can be obtained by applying $s_0s_4$ on the
first one in (\ref{bi_BT}). We also get the second equation in (\ref{bi_BT:1}) by applying
$s_0s_4s_6$ on the second one in (\ref{bi_BT}). Other equations can be derived in a similar
manner. By applying the translation
$T_{14}^{n'}\widehat{T}_{34}^m\widehat{T}_{40}^l\widehat{T}_{13}^k$ to the bilinear relations
(\ref{bi_BT:1}) and noticing (\ref{lattice-tau}), we get
\begin{equation}
\begin{array}{l}
(\alpha_0+\alpha_2+\alpha_4-m)
\,t^{-\frac{1}{2}}\tau_{k,l-1,m-1,n'}\tau_{k-1,l-1,m-1,n'+1}\\
\hskip20mm
+\tau_{k,l-1,m,n'+1}\tau_{k-1,l-1,m-2,n'}
+\tau_{k,l,m,n'}\tau_{k-1,l-2,m-2,n'+1}=0,\\[1mm]
(\alpha_0+\alpha_2+\alpha_4-m)\tau_{k-1,l-1,m-1,n'}\tau_{k,l-1,m-1,n'}\\
\hskip20mm
+\tau_{k,l-1,m,n'}\tau_{k-1,l-1,m-2,n'}-\tau_{k,l,m,n'}\tau_{k-1,l-2,m-2,n'}=0,
\end{array}
\end{equation}
and then (\ref{bi_n}) for the hypergeometric $\tau$ functions. Similarly, we obtain for the
hypergeometric $\tau$ functions (\ref{bi_m}), (\ref{bi_m'}) and (\ref{bi_BT'}) from (\ref{bi_BT:2}),
(\ref{bi_BT:3}) and (\ref{bi_BT:4}), respectively. The constraints
\begin{equation}
f_0=f_4-t,\quad f_3=f_4-1,
\end{equation}
yield
\begin{equation}
\begin{array}{l}
\tau_0s_4s_2s_0(\tau_0)
=s_4(\tau_4)s_2s_4(\tau_4)-t\tau_1s_4s_2s_1(\tau_1),\\[1mm]
\tau_0s_1s_2s_0(\tau_0)
=\tau_4s_1s_2s_4(\tau_4)-ts_1(\tau_1)s_2s_1(\tau_1),\\[1mm]
\tau_3s_4s_2s_3(\tau_3)
=s_4(\tau_4)s_2s_4(\tau_4)-\tau_1s_4s_2s_1(\tau_1),
\end{array}
\end{equation}
and 
\begin{equation}
\begin{array}{l}
\tau_3s_3(\tau_3)-\tau_0s_0(\tau_0)=(t-1)\tau_1s_1(\tau_1),\\[1mm]
t\tau_3s_3(\tau_3)-\tau_0s_0(\tau_0)=(t-1)\tau_4s_4(\tau_4),
\end{array}
\end{equation}
from which we obtain (\ref{bi_constraint1}) and (\ref{bi_constraint2}), respectively. Due to (\ref{def_h}) we have the relation
\begin{equation}
h_0-h_3=(t-1)\left[f_2f_4+\dfrac{1}{2}(1-\alpha_3-\alpha_4)\right].
\end{equation}
Then we get the bilinear relations 
\begin{equation}
\begin{array}{l}
D\,\tau_0\cdot\tau_3
=t^{\frac{1}{2}}s_4(\tau_4)s_2s_1(\tau_1)
+\dfrac{1}{2}(1-\alpha_3-\alpha_4)\tau_0\tau_3,\\[3mm]
D\,\tau_0\cdot\tau_3
=t^{\frac{1}{2}}\tau_4s_4s_2s_1(\tau_1)
+\dfrac{1}{2}(1-\alpha_3+\alpha_4)\tau_0\tau_3,\\[3mm]
D\,\tau_0\cdot\tau_3
=t^{\frac{1}{2}}\tau_1s_1s_2s_4(\tau_4)+\dfrac{1}{2}(\alpha_0-\alpha_1)\tau_0\tau_3,
\end{array}   \label{bi_D}
\end{equation}
where $D$ denotes Hirota's differential operator defined by $D\,g\cdot
f=t\left(\dfrac{dg}{dt}f-g\dfrac{df}{dt}\right)$. By applying the translation
$T_{14}^{n'}\widehat{T}_{34}^m\widehat{T}_{40}^l\widehat{T}_{13}^k$ to the first bilinear relation
of (\ref{bi_D}), one gets
\begin{equation}
\begin{array}{l}
\left[D+\dfrac{1}{2}\left(\alpha_3+\alpha_4-k+l+n'-\dfrac{1}{2}\right)\right]\,
\tau_{k,l,m,n'}\cdot\tau_{k,l-1,m-1,n'}
=-t^{\frac{1}{2}}(t-1)^{-\frac{1}{2}}
\tau_{k-1,l-1,m-1,n'+1}\tau_{k+1,l,m,n'-1},
\end{array}
\end{equation}
which is reduced to the first relation of (\ref{bi_D'}). The second and third relations of
(\ref{bi_D}) also yield their counterparts in (\ref{bi_D'}).


\begin{thebibliography}{00}
%
\bibitem{AS}
 Handbook of mathematical functions with formulas, graphs, and mathematical tables, 
 Edited by M. Abramowitz and I. A. Stegun. 
 Dover Publications, Inc., New York, 1992. 
%
%
\bibitem{Agafonov_1}
 Agafonov, S. I. 
 ``Imbedded circle patterns with the combinatorics of the square grid and discrete Painlev\'e equations.''
 \textit{Discrete Comput. Geom.} 29, no. 2 (2003): 305--319.
%
\bibitem{Agafonov_2}
 Agafonov, S. I. 
 ``Discrete Riccati equation, hypergeometric functions and circle patterns of Schramm type.'' 
 \textit{Glasg. Math. J.} 47, no. A (2005): 1--16.
%
\bibitem{Agafonov-Bobenko:2000}
 Agafonov, S. I., and Bobenko, A. I. 
 ``Discrete $Z^\gamma$ and Painlev\'e equations.''
 \textit{Internat. Math. Res. Notices} 2000, no. 4: 165--193.
%
\bibitem{Agafonov-Bobenko:2003}
 Agafonov, S. I., and Bobenko, A. I. 
 ``Hexagonal circle patterns with constant intersection angles and discrete Painlev\'e and Riccati equations.''
 \textit{J. Math. Phys.} 44, no. 8 (2003): 3455--3469.
%
\bibitem{Beardon-Stephenson} 
 Beardon, A. F., and Stephenson, K. 
 ``The uniformization theorem for circle packing.''
 \textit{Indiana Univ. Math. J.} 39, no. 4 (1990): 1383--1425.
%
\bibitem{Bobenko} 
 Bobenko, A. I. 
 "Discrete conformal maps and surfaces" in 
 Symmetries and integrability of difference equations (Canterbury, 1996), 97--108, 
 London Math. Soc. Lecture Note Ser. 255
 (Cambridge Univ. Press, Cambridge, 1999). 
%
\bibitem{Bobenko-Hoffmann}
 Bobenko, A. I., and Hoffmann, T. 
 ``Hexagonal circle patterns and integrable systems: patterns with constant angles.''
 \textit{Duke Math. J.} 116, no. 3 (2003): 525--566.
%
\bibitem{Bobenko-Hoffmann-Suris}
 Bobenko, A. I., Hoffmann, T., and Suris, Y. B. 
 ``Hexagonal circle patterns and integrable systems: patterns with the multi-ratio property and Lax equations on the regular triangular lattice.''
 \textit{Int. Math. Res. Not.} 2002, no. 3: 111--164.
%
\bibitem{BP_1}
 Bobenko, A. I., and Pinkall, U. 
 ``Discrete isothermic surfaces.''
 \textit{J. Reine Angew. Math.} 475 (1996): 187--208. 
%
\bibitem{BP_2}
 Bobenko, A. I., and Pinkall, U. 
 ``Discretization of surfaces and integrable systems.''
 Discrete integrable geometry and physics (Vienna, 1996), 3--58, 
 Oxford Lecture Ser. Math. Appl., 16, Oxford Univ. Press, New York, 1999. 
%
\bibitem{HKM}
 Hay, M., Kajiwara, K., and Masuda, T.
 ``Bilinearization and special solutions to the discrete Schwarzian KdV equation.''
 \textit{J. Math-for-Ind.} 3, (2011): 53--62.
%
\bibitem{Masuda}
 Masuda, T. 
 ``Classical transcendental solutions of the Painlev\'e equations and their degeneration.''
 \textit{Tohoku Math. J.} 56, no. 4 (2004): 467--490.
%
\bibitem{NRGO}
 Nijhoff, F. W., Ramani, A., Grammaticos, B., and Ohta, Y. 
 ``On discrete Painlev\'e equations associated with the lattice KdV systems and the Painlev\'e VI equation.''
 \textit{Stud. Appl. Math.} 106, no. 3 (2001): 261--314. 
%
\bibitem{Rodin}
 Rodin, B. 
 ``Schwarz's lemma for circle packings.''
 \textit{Invent. Math.} 89, no. 2 (1987): 271--289.
%
\bibitem{Schramm} 
 Schramm, O. 
 ``Circle patterns with the combinatorics of the square grid.''
 \textit{Duke Math. J.} 86, no. 2 (1997): 347--389.
%
\bibitem{Stephenson:book} 
 Stephenson, K. 
 \textit{Introduction to circle packing}, 
 New York: Cambridge University Press, 2005.
%
\bibitem{Thurston}
 Thurston, W. P. 
 ``The finite Riemann mapping theorem.''
 Invited address, International Symposium in Celebration of the Proof of the Bieberbach Conjecture (Purdue University, 1985). 
\end{thebibliography}
\end{document}